# Environmental Policy in General Equilibrium
# under Market Power and Price Discrimination

Tengjiao Chen
Independent Researcher
Vienna, VA 22182

Daniel H. Karney*
Ohio University
Department of Economics
Athens, OH 45701

January 2025

**Abstract.** This study constructs a novel analytical general equilibrium model to compare environmental policies in a setting where oligopolistic energy firms engage in third-degree price discrimination across residential consumers and industrial firms. Closed-form solutions demonstrate the impact on prices and quantities. The resulting welfare change is decomposed across three distortions: output, price discrimination, and externality. This study finds that the output distortion and price discrimination welfare effects generally move in opposite directions under policies such as an emission tax or a two-part instrument. Numerical analysis compares policies and finds scenarios where the output distortion and price discrimination welfare changes fully offset and thus leaves the net welfare gain of the externality correction. In this way, environmental policy can be designed to mitigate output distortion welfare concerns when firms have market power.

*Corresponding author; email address: karney@ohio.edu. We appreciate the helpful comments from Glenn Dutcher, Eyal Frank, Don Fullerton, Madhu Khanna, Erica Meyers, David Saal, and from the seminar participants at Kent State University, Loughborough University, Ohio University, University of Illinois at Urbana-Champaign, and University of Southampton. Neither author has a financial disclosure or conflict of interest to declare.

# 1. Introduction

Imposing a Pigouvian tax on a polluting monopolist leads to ambiguous welfare effects as the firm further contracts output (Buchanan, 1969). The first-best response employs two policy instruments, a tax to correct the externality and a subsidy to correct the output distortion, although an environmental regulator may only have authority to address the externality leading to second-best approaches (Cropper and Oates, 1992). The energy sector is a focus of environmental regulation since the combustion of fossil fuels contributes the majority share of U.S. greenhouse gas (GHG) emissions (EPA, 2022). However, energy sector firms have market power due to natural monopolies, barriers to entry, and government regulations (Posner, 1969; Baumol, 1977; Borenstein and Bushnell, 2015).[1] Thus, energy sector firms having market power leads to the concern that environmental policy to reduce GHG emissions may further distort output, and so reduces the net welfare gain from the externality correction envisaged by an environmental policy.

The energy sector market structure and output characteristics also enable price discrimination whereby homogeneous goods such as electricity and natural gas are sold to different market segments at different prices leading to another distortion. Kahn-Lang (2022) finds evidence of third-degree price discrimination within deregulated U.S. residential electricity markets. Globally, price discrimination is a significant concern in deregulated electricity markets (Simshauser, 2018; Waddams Price, 2018). In regulated energy markets, regulatory capture can potentially lead to differentiated prices across market segments too (Laffont and Tirole, 1993).[2]

There are differentiated prices across segments in the United States energy markets. The U.S. Energy Information Agency (EIA) reports that during 2019 the average residential consumer paid approximately 90 percent more for electricity than the average industrial user (EIA, 2021). A similar pattern arises in the natural gas market as the average U.S. residential gas price in 2019 was nearly 40 percent higher than the average commercial user price (EIA, 2022). Our study's analytical setup focuses on the price gap between what residential users pay for energy products and what the industrial sector pays for the same homogeneous goods. Our results are consistent with the price differences being driven by price discrimination after accounting for the additional distribution cost of providing energy services to the residential sector.

---

[1] In the global oil market, Asker et al. (2023) finds that OPEC's market power in the oil market reduced welfare by nearly $5 trillion, but Asker et al. (2024) then finds nearly the same magnitude increase in welfare from avoided greenhouse gas (GHG) emissions due to the cartel's output restriction. In the United States, Mansur (2007) finds empirical evidence that firms with market power in the electric power sector increase prices by distorting output decisions to increase profits but emissions coincidentally fell too. These studies add to the empirical literature finding market power in a variety of energy markets (Wolfram, 1998; Wolfram, 1999; Borenstein *et al.*, 2002; Puller, 2007).

[2] Segmenting is a necessary condition for third-degree price discrimination. However, firms can segment without third-degree price discriminating; for example, a firm could have different marketing campaigns to different segments but still charge the same price. Conversely, a firm charging the same price to different segments does not mean there is not third-degree price discriminating due to potentially different segment elasticities or marginal costs.

The aggregate U.S. energy sector – comprising the electricity, natural gas, and petroleum sectors – is relatively large and accounts for approximately 6 percent of 2019 U.S. GDP (EIA, 2020), and this implies environmental policies to reduce GHG emissions from the combustion of fossil fuels affects general equilibrium prices and quantities. Thus, this study compares the general equilibrium welfare effects of different environmental policies to reduce energy sector emissions given market power and price discrimination. Theoretical and numerical results show that environmental policies can mitigate welfare concerns about additional output distortions given market power in this setting while addressing the negative externality of GHG emissions.

A large literature since Pigou (1932) examines the regulation of environmental externalities in many applied theory contexts including in partial equilibrium under oligopoly (e.g., Levin, 1985) and price discrimination (e.g., Laffont and Tirole, 1996), as well as in general equilibrium (e.g., Bhattacharyya, 1996). Our study combines all three features in a single analytical general equilibrium model by deriving and employing a new modeling technique. Additional studies investigate environmental policies and their effects under different market structures in partial equilibrium (Smith, 1976; Baumol and Oates, 1988; Misiolek, 1988; Conrad and Wang, 1993; Ebert and von dem Hagen, 1998).[3] However, partial equilibrium models cannot analyze the effects on factor prices or intermediate goods markets, and thus a general equilibrium approach is necessary in the context of environmental policy affecting the entire energy sector. Other studies investigate tax incidence and welfare consequences using general equilibrium models. Browning (1997) shows interactions between labor taxes and welfare given imperfect competition. Fullerton and Metcalf (2002) uses a general equilibrium model to evaluate the welfare impact of implementing a cap-and-trade program with a pre-existing labor tax distortion, and that study finds that the existence of monopoly reduces the welfare gain from environmental policy.[4] However, to the best of our knowledge, the prior literature does not evaluate the welfare impact of price discrimination in general equilibrium with respect to environmental policy.

---

[3] Asch and Seneca (1976) and Misiolek (1980) derive the optimal tax rate to achieve socially efficient output level given a monopoly firm and externalities in the case of linear demand and cost. Barnett (1980) uncovers a general Pigouvian tax rule under monopoly suggesting that corrective taxes may reduce social costs due to externality correction but also could increase welfare loss from the underproduction of the final products.

The model in this study builds on the general equilibrium models in Fullerton and Metcalf (2002) and Baylis *et al.* (2014), but we innovate by modeling an oligopolistic energy sector with third-degree price discrimination across residential consumers and industrial firms. The model can specialize to monopoly or perfect competition via the number of equivalent oligopolistic firms (alternatively interpreted as a competition index) in the energy sector.

This study contributes to the literature in five ways. First, it develops a new modeling technique to embed an oligopolistic production sector with price discrimination within an analytical general equilibrium framework and finds solutions. Furthermore, this oligopoly sector can be specialized to either a pure monopoly or perfect competition, and thus allows one general equilibrium model to consider the full range of market structures. The prior analytical general equilibrium literature assumes the energy (or pollution-producing) production sector conforms to either perfect competition or monopoly, and no prior model allows for price discrimination. This modeling innovation can be used in settings other than the analysis of environmental policy.

Second, this study derives a welfare formula that separately identifies the output distortion and price discrimination effects in addition to the standard externality effect. The expression verifies that the difference in prices without price discrimination is only due to differential taxes across sectors and distribution costs, although the output distortion remains as well as the distortion from a sub-optimally regulated externality as previously shown in the literature. Prior analytical general equilibrium models do not simultaneously consider this combination of features and thus this study formally identifies these three distortion terms in the same welfare formula.

Third, using the welfare formula, this study identifies conditions when the output distortion and price discrimination effects have opposite signs when tightening environmental regulation. The resulting theorems show necessary and sufficient conditions for offsetting output distortion and price discrimination welfare effects. If the output distortion and price discrimination effects fully offset, then the net welfare effect remaining is the environmental externality correction. These scenarios address the concern that environmental policy in an imperfectly competitive market further exacerbates the existing output distortion and potentially reduces welfare.

---

[4] Other researchers derive optimal tax rules with dynamic models. For example, Chang *et al.* (2009) proposes a socially optimal tax policy approach for externalities under market imperfections with specific functional forms and Golosov *et al.* (2014) identifies an optimal-tax formula using dynamic stochastic general equilibrium model with an externality.

Fourth, numerical analysis compares environmental policies such as a standard emission tax and a two-part instrument in this unexplored setting. The emission tax and two-part instrument are normalized to generate the same total emission reduction, but the two-part instrument yields a higher net welfare as the output distortion and price discrimination effects offset. Also, the emission tax case finds a positive change in the oligopolistic energy firms' profits while the revenue recycling and two-part instrument scenarios find a negative change. Thus, energy firms prefer an emission tax relative to the two-part instrument given this study's parameterization as the two-part instrument reduces profits from price discrimination.

Fifth, numerical analysis also demonstrates that observed energy prices constrain the set of potential assumptions for unobserved parameters governing the oligopolistic energy sector. Specifically, the observed industrial energy price determines the industrial sector's energy input demand elasticity given an assumed value for the residential consumers' price elasticity of demand and observed energy price. Thus, this exercise cautions that not all empirical estimates of energy demand elasticities in the literature are necessarily compatible with the model structure given observed prices. Finally, we show how numerical results change if the researcher exogenously imposes a perfect competition or monopoly market structure.

The paper proceeds as follows. Section 2 presents the model setup along with the log-linear equations. Section 3 derives the welfare formula and related theorems. Section 4 provides the closed-form solutions. Section 5 conducts the numerical analysis. Section 6 concludes.

# 2. Model

This section begins with an exposition of the model and concludes with the system of log-linear equations describing how the endogenous variables respond to changes in an emission tax.

## 2.A Setup

The general equilibrium model contains an oligopolistic energy sector, a competitive industrial sector, and many identical residential consumers. The energy sector has $n$ identical model firms employing capital $(K_E)$ and emissions $(Z)$ to produce energy $(E)$ via a constant returns to scale (CRS) production function $E = E(K_E, Z)$ and using the convention that emissions are moved to the input side of the production function.[5] Arocena *et al.* (2012) provides empirical evidence of constant returns to scale in the energy sector despite the existence of market power.

In this model, $n$ can vary from $n = 1$ (monopoly) to $n = 2$ (duopoly) through $n \to \infty$ (perfect competition). The number $n$ can more generally be interpreted as a measure of competition, or competition index, and not necessarily the actual number of firms in the energy market. The homogenous energy good is sold via third-degree price discrimination to the industrial sector and residential consumers. The industrial sector pays price $p_{EX}$ for energy while the residential price is $p_{ER}$, generally, $p_{EX} < p_{ER}$. The energy firms incur a pre-existing, positive per unit tax on emissions $(t_Z > 0)$ but due to market power secures rents and thus earn economic profits $(\Pi_E > 0)$.

The competitive industrial sector uses capital $(K_X)$ and energy $(E_X)$ to produce a final good $(X)$ via the CRS production function $X = X(K_X, E_X)$. The output is sold to the residential consumers at price $p_X$. The per unit energy cost paid by industrial firms is $p_{EX}$, but $t_{EX}$ is a per unit commodity tax on energy products sold to the industrial sector such that the oligopolistic energy firms receive $(p_{EX} - t_{EX})$ for every unit of output sold to the industrial sector. Similarly, the energy price paid by residential consumers is $p_{ER}$, but the energy sector receives $(p_{ER} - \delta - t_{ER})$, where $t_{ER}$ is a per unit tax on energy products sold to the residents and $\delta$ is the additional distribution cost to supply the residential sector.[6]

The many identical consumers own a fixed total amount of "capital", denoted by $\overline{K}$, which can be interpreted more generally as a composite of clean capital and labor. The capital resource constraint becomes $K_E + K_X = \overline{K}$, where $K_E$ and $K_X$ denote the amount of capital employed in energy and industrial sectors, respectively, as the clean inputs to production. Since capital is not sector specific and thus perfectly mobile, then the gross, pre-tax price of capital $(q_K)$ is the same in both energy and industrial sectors. However, the model includes (potential) sector specific factor taxes such that the net, after-tax price of the capital in the energy and industrial sectors are given by $p_{KE} = q_K + t_{KE}$ and $p_{KX} = q_K + t_{KX}$, where the per unit capital taxes are $t_{KE}$ on $K_E$ and $t_{KX}$ on $K_X$, respectively.

---

[5] Emissions $(Z)$ can model a single externality – for example, greenhouse gas emissions as shown in the numerical exercise below in Section 5 – or $Z$ could represent an aggregate function of many externalities.

[6] This additional cost to energy firms selling to the residential can arise due to a variety of factors including the need to build small-capacity infrastructure, extra customer service burdens, and regulatory requirements. The assumption of identical energy firms in the model also abstracts from the possibility that some firms may only sell to the industrial sector instead of both industrial and residential users (or perhaps only the residential sector), but the aggregate behavior of the energy sector is not affected by this abstraction due to the assumption of constant returns to scale production.

The residents consume energy products $(E_R)$ directly as a final good as well as the output $(X)$ from the industrial sector. The utility function of the representative consumer is given by $U(X, E_R; Z)$ and it is assumed that $\partial U/\partial X > 0$, $\partial U/\partial E_R > 0$, and $\partial U/\partial Z < 0$. The level of emissions is taken as exogenous in utility maximization problem by the atomistic residential consumer and thus $Z$ is offset by a semicolon in the utility function. All net tax revenue is returned to consumer via a lump-sum rebate $(T)$ and thus the budget constraint becomes $I \equiv q_K \overline{K} + \delta E_R + \Pi_E + T = p_X X + p_{ER} E_R$, where energy sector profits are given by $\Pi_E = (p_{ER} - \delta - t_{ER} - \gamma)E_R + (p_{EX} - t_{EX} - \gamma)E_X$ and where $\gamma$ is the marginal cost of energy production.[7]

In this single-period, long-run model, all energy produced is consumed according to the supply-demand condition $E = E_X + E_R$. Following Fullerton and Metcalf (2002) and the related literature, this model assumes perfect certainty, no transactions costs, and availability of lump-sum transfers.

## 2.B Log-Linear System

The model is log-linearized to analyze how small changes in the exogenous taxes – such as increasing the emission tax or subsidizing the clean input to energy production – affect the equilibrium outcomes given the level energy sector competitiveness.[8]

To start, totally differentiate the capital resource constraint to find:

$$\omega_E \widehat{K}_E + (1 - \omega_E)\widehat{K}_X = 0, \tag{1}$$

---

[7] The budget constraint implicitly assumes an existing, regulated (or publicly owned) distribution network and thus the distribution charges along with net profits are returned to the resident sector as both the public and the shareholders. Also, the total distribution cost $(\delta E_R)$ must be included in income for total expenditures to equal total income as required in general equilibrium.

[8] The log-linear version of the model provides the changes in prices and quantities from one long-run equilibrium to another long-run equilibrium because of a policy change but does not analyze the transition path.

where $\omega_E \equiv K_E / \overline{K}$ is the share of total capital employed in the energy sector. Appendix A catalogs all parameter definitions. The "hat" over a variable indicates a proportional change (e.g., $\hat{R}_E = dK_E / K_E$). Mechanically, a capital use increase in the energy sector must lead to a proportional decrease in the industrial sector, and vice versa.

Next, define $\sigma_U > 0$ as the elasticity of substitution between final goods $X$ and $E_R$ to reflect the representative consumer's preferences. Then, totally differentiate the definition of $\sigma_U$ finds:

$$\hat{X} - \hat{E}_R = \sigma_U (\hat{p}_{ER} - \hat{p}_X), \tag{2}$$

such that an increase in the relative price of energy to the residents, $(\hat{p}_{ER} - \hat{p}_X) > 0$, leads to an increase in the relative consumption of industrial output compared to energy, $(\hat{X} - \hat{E}_R) > 0$, all else equal.

Continuing, let the elasticity of substitution in the industrial sector be denoted as $\sigma_X > 0$. Then, totally differentiating the definition of $\sigma_X$ yields:

$$\hat{E}_X - \hat{R}_X = \sigma_X (\hat{p}_{KX} - \hat{p}_{EX}). \tag{3}$$

Firms in the industrial sector maximize their profits but earn zero profits because of competition and CRS production. Totally differentiating the production function and substituting the first order conditions (FOCs) of the profit maximization problem reveals:

$$\hat{X} = (1 - \theta_E^X)\hat{R}_X + \theta_E^X \hat{E}_X, \tag{4}$$

where $\theta_E^X \equiv p_{EX} E_X / p_X X$ is the share of total revenue of the industrial firms that is spending on energy input $E_X$. The zero-profit condition is given by $p_X X = p_{KX} K_X + p_{EX} E_X$ and then totally differentiating leads to:

$$\hat{X} + \hat{p}_X = (1 - \theta_E^X)(\hat{R}_X + \hat{p}_{KX}) + \theta_E^X (\hat{E}_X + \hat{p}_{EX}). \tag{5}$$

Recall the oligopolistic energy firms produce a homogenous good $(E)$ sold to industrial firms as an intermediate good $(E_X)$ or residential consumers as a final good $(E_R)$. The market clearing condition is $E = E_X + E_R$ and totally differentiating yields:

$$\hat{E} = \varphi_X \hat{E}_X + \varphi_R \hat{E}_R, \tag{6}$$

where $\varphi_X \equiv E_X/E$ is the share of total energy products sold to the industrial sector. Similarly, $\varphi_R \equiv E_R/E$ as the share of total energy products sold to the residents, and $\varphi_X + \varphi_R = 1$.

Next, energy is produced via the two-input production function $E = E(K_E, Z)$ and let $\sigma_E > 0$ denote the elasticity of substitution between inputs. Totally differentiating the definition of $\sigma_E$ provides the log-linear equation:

$$\hat{Z} - \hat{K}_E = \sigma_E(\hat{p}_{KE} - \hat{t}_Z), \tag{7}$$

and define the proportional change in the emission tax as $\hat{t}_Z \equiv dt_Z/t_Z$ because the model assumes an existing tax ($t_Z > 0$) levied on emissions.

Here, the model departs from the standard methodology by adding oligopolistic firm behavior. Furthermore, this new modeling technique is flexible with respect to the number of firms such that both pure monopoly and perfect competition are special cases. The model assumes quantity setting via Cournot competition. Puller (2007) finds empirical evidence that electricity sector firms engage in a market structure with outcomes that closely match Cournot model predictions. Also, the Cournot model is equivalent to a two-stage oligopoly model whereby firms choose production capacity at the first stage and then engage in price competition via Bertrand in the second stage (Kreps and Scheinkman, 1983). The two-stage interpretation fits many energy sector settings such as fixed capacity power plants and oil refineries that sell electricity and gasoline, respectively.

As Appendix B derives, the oligopolistic energy sector output price changes are a function of the number of firms, or competition index, ($n$) and are found:

$$\hat{p}_{EX} = \left(\frac{n\varepsilon_{EX}}{1 + n\varepsilon_{EX}}\right)\left(\frac{t_{EX}}{p_{EX}}\hat{t}_{EX} + \frac{\gamma}{p_{EX}}\hat{\gamma}\right), \tag{8}$$

and:

$$\hat{p}_{ER} = \left(\frac{n\varepsilon_{ER}}{1 + n\varepsilon_{ER}}\right)\left(\frac{t_{ER}}{p_{ER}}\hat{t}_{ER} + \frac{\gamma}{p_{ER}}\hat{\gamma}\right), \tag{9}$$

recalling $\gamma$ is the marginal cost of production, and $\hat{\gamma} = d\gamma/\gamma$, $\hat{t}_{EX} = dt_{EX}/t_{EX}$ and, $\hat{t}_{ER} = dt_{ER}/t_{ER}$. The marginal cost in this long-run model includes the implicit marginal cost of new production capacity as well as the cost other inputs such as labor since $K$ is defined to be a composite of all clean inputs including land, labor, and capital. Also, define $\varepsilon_{EX} \equiv (\partial E_X/E_X)/(\partial p_{EX}/p_{EX})$ as industry $X$'s uncompensated price elasticity of demand for the energy input $E_X$, and similarly $\varepsilon_{ER} \equiv (\partial E_R/E_R)/(\partial p_{ER}/p_{ER})$ is the residential uncompensated price elasticity of demand for energy. As Appendix B shows, it must hold that $\varepsilon_{EX}, \varepsilon_{ER} < -1/n$ at the prevailing prices and quantities, and, again, if $n = 1$ then the energy sector is served by a monopolist, but as $n \to \infty$ the sector tends towards perfectly competitive outcomes. The value for $n$ is fixed in the log-linearization meaning small tax policy changes do not affect the competitiveness of the energy sector.[9]

Equations (8) and (9) come from the energy sector's FOCs given by $n(p_{EX} - t_{EX}) + \frac{p_{EX}}{\varepsilon_{EX}} = n\gamma$ and $n(p_{ER} - \delta - t_{ER}) + \frac{p_{ER}}{\varepsilon_{ER}} = n\gamma$, but $\delta$ drops out of the log-linear equations since it is fixed. As discussed in Section 5, these FOCs govern the relationship between the observable prices and the unobservable parameters such as marginal cost and the demand elasticities. Also, the parameters $\sigma_U$ and $\varepsilon_{ER}$ are not independent and governed by the equation $\varepsilon_{ER} = -\theta_E^R - (1 - \theta_E^R)\sigma_U$, where $\theta_E^R \equiv p_{ER}E_R/I$ is the share of total income residents spend on energy.[10] This relationship implies that $\varepsilon_{ER} < 0$ when $\sigma_U > 0$, whereby the elasticity of substitution is positive while the own-price demand elasticity is negative. Similarly, it can be shown that $\varepsilon_{EX} = -\theta_E^X - (1 - \theta_E^X)\sigma_X$ with $\sigma_X > 0$ and $\varepsilon_{EX} < 0$, where $\theta_E^X \equiv p_{EX}E_X/p_X X$ is the share of total revenue industrial firms spending on energy.

Continuing, the oligopolistic energy sector output changes according to the equation derived from its production function $E = E(K_E, Z)$ and given by:

$$\hat{E} = \rho_K^E \hat{K}_E + \rho_Z^E \hat{Z}, \tag{10}$$

where $\rho_Z^E \equiv \frac{t_Z Z}{\gamma E}$ is the ratio of total emission taxes to total marginal energy production costs and, similarly, $\rho_K^E \equiv \frac{p_{KE} K_E}{\gamma E}$ is the ratio of total capital costs to total marginal energy production cost. Equation (10) differs from equation (4) in that it does not contain the output price. Also, the zero-profits condition does not apply since the energy firms have market power. However, the constant returns to scale assumption implies that total production costs equal the input costs, and it follows that:

$$\hat{\gamma} + \hat{E} = \rho_K^E(\hat{p}_{KE} + \hat{K}_E) + \rho_Z^E(\hat{t}_Z + \hat{Z}). \tag{11}$$

Finally, recall the capital tax relationships, $p_{KE} = q_K + t_{KE}$ and $p_{KX} = q_K + t_{KX}$, such that total differentiation yields:

$$\hat{p}_{KE} = \beta_E \hat{q}_K + \hat{t}_{KE}, \tag{12}$$

$$\hat{p}_{KX} = \beta_X \hat{q}_K + \hat{t}_{KX}, \tag{13}$$

where $\beta_E \equiv q_K/p_{KE}$ and $\beta_X \equiv q_K/p_{KX}$. Also, define $\hat{t}_{KE} \equiv dt_{KE}/p_{KE}$ and $\hat{t}_{KX} \equiv dt_{KX}/p_{KX}$, so that both pre-existing capital taxes can be set to zero, if necessary. The system is closed by setting capital as numeraire. Specifically, set $q_K = 1$ such that:

$$\hat{q}_K = 0, \tag{14}$$

and thus equations (12) and (13) simplify to $\hat{p}_{KE} = \hat{t}_{KE}$ and $\hat{p}_{KX} = \hat{t}_{KX}$, respectively. The alterative specification for $\hat{t}_{KE}$ and $\hat{t}_{KX}$, along with the numeraire definition, enables the elimination of parameters $\beta_E$ and $\beta_X$ from the closed-from solutions.

Equations (1)-(14) define a system of linear equations in 14 endogenous variables $\{\hat{K}_E, \hat{Z}, \hat{E}, \hat{E}_R, \hat{E}_X, \hat{K}_X, \hat{X}, \hat{p}_{ER}, \hat{p}_X, \hat{p}_{EX}, \hat{p}_{KX}, \hat{p}_{KE}, \hat{\gamma}, \hat{q}_K\}$ and 5 potential policy variables given by the exogenous taxes $\{\hat{t}_Z, \hat{t}_{ER}, \hat{t}_{EX}, \hat{t}_{KE}, \hat{t}_{KX}\}$. For example, one can assess effects on the endogenous variables from a policy change increasing the emission tax ($\hat{t}_Z > 0$) holding the other policies constant ($\hat{t}_{ER} = \hat{t}_{EX} = \hat{t}_{KE} = \hat{t}_{KX} = 0$). Other policies can be assessed too such as combining taxes and subsidies to address multiple distortions under a balanced-budget condition. The model generates other variables of potential interest including change in the lump-sum rebate ($\hat{T}$) and energy firm profits ($\hat{\Pi}_E$) as shown in Appendix C.

# 3. Welfare

This section derives welfare change formulae under imperfect competition, price discrimination, and a negative externality. To start, totally differentiate the representative consumer's utility function and substitute in the FOCs from the utility maximization problem to find (see Appendix D):

$$dU = \lambda p_X dX + \lambda p_{ER} dE_R + \frac{\partial U}{\partial Z} dZ$$

where $\lambda$ is the Lagrange multiplier on the budget constraint. Then, define $\mu \equiv -(\partial U/\partial Z)/\lambda > 0$ as the monetary equivalent utility value of externality, also known as the marginal environmental damage (MED). The MED equals the social cost of carbon (SCC) when considering carbon emissions exclusively.

Appendix D further derives equations (15) and (16) in the Harberger style where the left-hand side is a monetary equivalent. Assuming $p_{KX} = p_{KE}$ (implicitly, $t_{KE} = t_{KX} \geq 0$) for now, the change in welfare as a function of the proportional changes in energy $(\hat{E}_X, \hat{E}_R)$ and emissions $(\hat{Z})$ is found:

$$\frac{dU}{\lambda I} = \underbrace{\left[ (p_{EX} - \gamma) \frac{E_X}{I} \hat{E}_X + (p_{ER} - \gamma) \frac{E_R}{I} \hat{E}_R \right]}_{\text{Market Power}} + \underbrace{(t_Z - \mu) \frac{Z}{I} \hat{Z}}_{\text{Externality}}, \qquad (15)$$

where equation (15) contains two distortions. The first distortion in brackets is an aggregate market power effect when price does not equal marginal cost. For instance, if the industrial sector's energy purchase price is more than marginal cost of production $(p_{EX} > \gamma)$, then price is above the competitive outcome and welfare increases when $\hat{E}_X > 0$, all else equal, as too few transactions occur under oligopoly. The second distortion is the externality effect. For this effect, welfare improves when the emission level falls $(\hat{Z} < 0)$ given the prevailing emission tax is less than the marginal damage from emissions $(t_Z < \mu)$. Furthermore, this term shows that emissions can be overtaxed $(t_Z > \mu)$ such that too much abatement occurs in the initial equilibrium compared to the social optimum. If all the distortion effects on the right-hand side are zero $((p_{EX} - \gamma) = (p_{EX} - \gamma) = (t_Z - \mu) = 0)$, then it is the first-best outcome.

The aggregate market power effect in equation (15) can be decomposed into an oligopoly output effect and a price discrimination effect as shown in equation (16). The oligopoly output effect arises because imperfect competition allows the oligopolistic firms to increase price and reduce output. A separate price discrimination effect occurs as the homogenous output is sold at different prices to residential consumers and the industrial sector. The three-effect form is found to be:

$$\frac{dU}{\lambda I} = \underbrace{(\varphi_X p_{EX} + \varphi_R p_{ER} - \gamma)\frac{E}{I}\hat{E}}_{\text{Oligopoly Output}} + \underbrace{(p_{ER} - p_{EX})\frac{E_R E_X}{EI}(\hat{E}_R - \hat{E}_X)}_{\text{Price Discrimination}} + \underbrace{(t_z - \mu)\frac{Z}{I}\hat{Z}}_{\text{Externality}}, \qquad (16)$$

where the difference between the weighted average of the energy prices $(\varphi_X p_{EX} + \varphi_R p_{ER})$ and the marginal cost of production $(\gamma)$ determines the size of the oligopoly output effect. To analyze the price discrimination effect, assume the residential price is higher than the industrial price $(p_{ER} > p_{EX})$, and thus a relative increase in residential use $(\hat{E}_R > \hat{E}_X)$ raises welfare as the energy buyer with the higher willingness to pay consumes more, all else equal. The externality effect is unchanged.[11]

Theorems 1 and 2 below examine the interactions between the oligopoly output and price discrimination welfare effects for policies that decrease total energy sector output $(\hat{E} < 0)$, such as an increase in the emission tax. These theorems do not consider the welfare change from the emission externality as the welfare improves from reduction in pollution when the MED is greater than the prevailing tax rate.

Theorem 1 provides sufficient conditions where the oligopoly output effect is negative and so decreases welfare, but where the price discrimination effect is positive from adjusting quantities to help offset the welfare loss. The price discrimination welfare offset helps address the common concern that welfare gains from the externality correction are then lost to additional output distortions in the presence of market power.

---

[11] Consider the special case where price discrimination does not occur for either technical or legal reasons so the homogeneous good has the same price across buyers. If so, then $p_{ER} = p_{EX} = p_E$, and the price discrimination distortion effect is zero. Also, the oligopoly output effect in equation (16) then simplifies to $(p_E - \gamma)(E/I)\hat{E}$ and one can show the market power effect in equation (15) simplifies to the same expression when the residential and industrial buyers face the same energy price.

**Theorem 1.** Assume a decrease in total energy sector output $\left(\hat{E} < 0\right)$. If so, then the oligopoly output effect is negative (as weighted-average energy price must be higher than its marginal cost). Furthermore, assume residential buyers pay a higher price for energy than industrial users ($p_{ER} > p_{EX}$). If so, then the price discrimination effect is positive when $\hat{E}_R > 0$ (implying $\hat{E}_X < 0$) or $\hat{E}_X < \hat{E}_R < 0$.

**Proof.** From equation (16), denote the oligopoly output effect as $\Psi \equiv (\varphi_X p_{EX} + \varphi_R p_{ER} - \gamma)\frac{E}{I}\hat{E}$ and the price discrimination effect as $\Omega \equiv (p_{ER} - p_{EX})\frac{E_R E_X}{EI}\left(\hat{E}_R - \hat{E}_X\right)$. By assumption $\hat{E} < 0$. Since energy production and income are positive values then $E/I > 0$. Market power implies $p_{ER} > \gamma$ and $p_{EX} > \gamma$, and recall $\varphi_R + \varphi_X = 1$. Thus, the weighted average of the two output prices is above the marginal cost so that $(\varphi_X p_{EX} + \varphi_R p_{ER} - \gamma) > 0$, and therefore $\Psi < 0$. Next, by assumption $p_{ER} > p_{EX}$, and $\frac{E_R E_X}{EI} > 0$ since all level quantities are positive. Then, $\hat{E}_R > 0$ and $\hat{E}_X < 0$, or $\hat{E}_X < \hat{E}_R < 0$, imply $\left(\hat{E}_R - \hat{E}_X\right) > 0$ and therefore $\Omega > 0$. **QED**

Next, Theorem 2 finds necessary and sufficient conditions for net welfare gains from reducing the price discrimination effect by outweighing additional losses from output distortions. Intuitively, even if total energy consumption decreases, total welfare can still increase if the more distorted market from price discrimination can sufficiently increase consumption. Theorem 2 shows that the ratio $(p_{EX} - \gamma)/(p_{ER} - \gamma)$ governs the sufficient condition for welfare improvement. A smaller ratio makes the sufficient condition easier to meet, all else equal, as a small ratio implies a large difference between $p_{ER}$ and $p_{EX}$, and thus implies significant welfare gains from correcting the price discrimination distortion are possible (recalling $p_{ER} > p_{EX}$).

**Theorem 2.** Assume a policy decreases total energy sector output $\left(\hat{E} < 0\right)$ and residential consumers face a higher initial energy price ($p_{ER} > p_{EX}$). Then, $\hat{E}_R > 0$ is a necessary condition to yield a net welfare gain with respect to the price discrimination and oligopoly output effects. Then, a sufficient condition to yield a net welfare gain with respect to the price discrimination and oligopoly output effects is given by:

$$\hat{E}_R > \frac{\varphi_X(p_{EX}-\gamma)}{\varphi_R(p_{ER}-\gamma)}\left(-\hat{E}_X\right) > 0, \text{ or equivalently, } dE_R > \frac{(p_{EX}-\gamma)}{(p_{ER}-\gamma)}\left(-dE_X\right).$$

**Proof.** Rewrite the expression $\Psi + \Omega > 0$ follows:

$$(\varphi_X p_{EX} + \varphi_R p_{ER} - \gamma)\frac{E}{I}\hat{E} + (p_{ER} - p_{EX})\frac{E_R}{E}\frac{E_X}{E}\frac{E}{I}(\hat{E}_R - \hat{E}_X) > 0$$

$$\Leftrightarrow (\varphi_X p_{EX} + \varphi_R p_{ER} - \gamma)\hat{E} + (p_{ER} - p_{EX})\varphi_R \varphi_X (\hat{E}_R - \hat{E}_X) > 0$$

$$\Leftrightarrow \varphi_X p_{EX}\hat{E} + \varphi_R p_{ER}\hat{E} - \gamma\hat{E} + \varphi_R \varphi_X p_{ER}\hat{E}_R - \varphi_R \varphi_X p_{EX}\hat{E}_R - \varphi_R \varphi_X p_{ER}\hat{E}_X + \varphi_R \varphi_X p_{EX}\hat{E}_X > 0$$

$$\Leftrightarrow \varphi_X \varphi_R p_{EX}\hat{E}_X + \varphi_R \varphi_R p_{ER}\hat{E}_R - \varphi_X \gamma \hat{E}_X - \varphi_R \gamma \hat{E}_R + \varphi_R \varphi_X p_{ER}\hat{E}_R + \varphi_R \varphi_X p_{EX}\hat{E}_X > 0$$

$$\Leftrightarrow \varphi_X \hat{E}_X(\varphi_X p_{EX} + \varphi_R p_{EX} - \gamma) + \varphi_R \hat{E}_R(\varphi_R p_{ER} + \varphi_X p_{ER} - \gamma) > 0$$

$$\Leftrightarrow \varphi_X \hat{E}_X(p_{EX} - \gamma) + \varphi_R \hat{E}_R(p_{ER} - \gamma) > 0.$$

By Theorem 1, $\Psi < 0$. Thus $\Psi + \Omega > 0$ implies $\Omega > 0$, and it must hold that $\hat{E}_R > \hat{E}_X$ when $p_{ER} > p_{EX}$. Assume $\hat{E}_R \leq 0$, then $\hat{E}_X < \hat{E}_R \leq 0$, but the above inequality does not hold and thus a contradiction. Therefore, $\hat{E}_R > 0$ is a necessary condition. Next, given $\hat{E} < 0$ and $\hat{E}_R > 0$, then $\hat{E}_X < 0$ (or $-\hat{E}_X > 0$) must hold by the market-clearing condition. Rewriting the above inequality yields the sufficient condition $\hat{E}_R > \frac{\varphi_X(p_{EX} - \gamma)}{\varphi_R(p_{ER} - \gamma)}\left(-\hat{E}_X\right) > 0$. **QED**

These theorems can be used to assess the potential welfare effects of policies that achieve the same emission reduction target. Fixing $\hat{Z}$ means the emission externality effect in equation (16) is constant and thus the relative net welfare change comes from the balance of the oligopoly output and price discrimination effects. The numerical exercises in Section 5 demonstrate a wide range of welfare outcomes given a fixed emission reduction target. Finally, Appendix D provides a six-term welfare equation that expands equation (16) under differential capital taxes ($t_{KE} \neq t_{KX}$) that introduces pre-existing distortions in the input market ($p_{KX} \neq p_{KE}$).

# 4. Analytical Results

This section presents the analytical results of the model presented in Section 2 where the system of log-linear equations (1)-(14) is solved to find the endogenous variables as a function of the policy variables. The numeraire definition ($\hat{q}_K = 0$) means $\hat{p}_{KE} = \hat{t}_{KE}$ and $\hat{p}_{KX} = \hat{t}_{KX}$. To build intuition for the results, analysis of the analytical solutions starts with the special case of only allowing the emission tax to vary and then turns to the general case where all taxes can vary. These closed-form and recursive solutions provided below show how market power mediates the effects of tax changes, such as a pollution tax, on the endogenous price and quantity variables in this general equilibrium setting.

## 4.A Special Case

Hold the non-emission tax policies constant relative to the numeraire (meaning $\hat{p}_{KE} = \hat{t}_{KE} = 0$ and $\hat{p}_{KX} = \hat{t}_{KX} = 0$ along with $\hat{t}_{ER} = \hat{t}_{EX} = 0$), and assume the emission tax exogenously increases ($\hat{t}_Z > 0$). Then, the closed-form solution for the marginal cost change becomes:

$$\hat{\gamma} = \rho_Z^E \hat{t}_Z > 0, \tag{17}$$

and the marginal cost of energy production ($\gamma$) always increases as mediated by the share of emission taxes relative to total energy production costs ($\rho_Z^E \equiv t_Z Z / \gamma E$). Since the cost share is less than one then marginal cost increases by a smaller percentage than the emission tax increase ($\hat{\gamma} < \hat{t}_Z$).

The energy prices always increase too relative to the numeraire and simply to:

$$\hat{p}_{EX} = \left(\frac{n\varepsilon_{EX}}{1 + n\varepsilon_{EX}}\right)\left[\frac{t_Z Z}{p_{EX} E}\right]\hat{t}_Z > 0, \tag{18}$$

$$\hat{p}_{ER} = \left(\frac{n\varepsilon_{ER}}{1 + n\varepsilon_{ER}}\right)\left[\frac{t_Z Z}{p_{ER} E}\right]\hat{t}_Z > 0, \tag{19}$$

where these price changes are mediated by modified cost share terms in the brackets. However, the market power effects in the parenthesis terms also impact the energy price changes. From Section 2, $\varepsilon_{EX}, \varepsilon_{ER} < -1/n$ with $n \geq 1$, and thus the terms in parenthesis are greater than one. Therefore, the energy price increases depend on the strength of the market power effects.

The final good price from the industrial sector is given by:

$$\hat{p}_X = \theta_E^X \hat{p}_{EX} = \theta_E^X \left(\frac{n\varepsilon_{EX}}{1 + n\varepsilon_{EX}}\right)\left[\frac{t_Z Z}{p_{EX} E}\right]\hat{t}_Z > 0, \tag{20}$$

where the industrial price is mediated by $\theta_E^X \equiv \frac{p_{EX} E_X}{p_X X}$, the share of total revenue of the industrial firms that is spending on input $E_X$. The price change for the final, industrial good is always smaller than the energy input price change for the industrial sector due to perfect competition ($\hat{p}_X / \hat{p}_{EX} = \theta_E^X < 1$).

The change in emissions is determined recursively according to:

$$\hat{Z} = -[\sigma_U(1 - \omega_E)\varphi_R](\hat{p}_{ER} - \hat{p}_X) - [\sigma_X(1 - \omega_E)(\varphi_X + \varphi_R\theta_E^X)]\hat{p}_{EX}$$
$$-[\sigma_E(1 - (1 - \omega_E)\rho_Z^E)]\hat{t}_Z, \tag{21}$$

where all the bracketed terms on the right-hand side of equation (21) are positive. Therefore, if $\hat{t}_Z > 0$ (implying $\hat{p}_{EX}, \hat{p}_{ER}, \hat{p}_X > 0$) and $(\hat{p}_{ER} - \hat{p}_X) > 0$, then emissions necessarily fall. Alternatively, one can rewrite (21) as follows:

$$\hat{Z} = [-\sigma_U(1 - \omega_E)\varphi_R\hat{p}_{ER} - \sigma_X(1 - \omega_E)\varphi_X\hat{p}_{EX} - \sigma_E(1 - (1 - \omega_E)\rho_Z^E)\hat{t}_Z]$$
$$-(\sigma_X - \sigma_U)(1 - \omega_E)\varphi_R\theta_E^X\hat{p}_{EX},$$

where the bracketed term is negative if $\hat{t}_Z > 0$ and thus emissions necessarily fall when $\sigma_X > \sigma_U$. In other words, emissions necessarily fall from an emission tax increase if the industrial sector substitutes more elastically in production than consumers in utility and our numerical analysis always finds $\sigma_X > \sigma_U$. Also, the converse does not hold as emissions do not necessarily increase if $\sigma_U > \sigma_X$.

## 4.B General Case

If all the taxes are allowed to vary, then the closed-form, general equilibrium solution for the change in the marginal cost of energy production is given as:

$$\hat{\gamma} = \rho_Z^E\hat{t}_Z + (1 - \rho_Z^E)\hat{t}_{KE}. \tag{22}$$

Here, the marginal cost of energy cost is positively related to the emission tax and energy sector capital tax, and thus $\hat{\gamma}$ can be negative given an emission tax increase if a prevailing capital tax sufficiently falls.

Next, equations (23) and (24) show the energy price changes as a function of multiple taxes. As with the specialized results above, the number of oligopolistic firms ($n$) and demand elasticities ($\varepsilon_{EX}, \varepsilon_{ER}$) mediate the energy prices changes:

$$\hat{p}_{EX} = \left(\frac{n\varepsilon_{EX}}{1 + n\varepsilon_{EX}}\right)\left[\frac{t_{EX}}{p_{EX}}\hat{t}_{EX} + \frac{p_{KE}K_E}{p_{EX}E}\hat{t}_{KE} + \frac{t_Z Z}{p_{EX}E}\hat{t}_Z\right], \tag{23}$$

$$\hat{p}_{ER} = \left(\frac{n\varepsilon_{ER}}{1 + n\varepsilon_{ER}}\right)\left[\frac{t_{ER}}{p_{ER}}\hat{t}_{ER} + \frac{p_{KE}K_E}{p_{ER}E}\hat{t}_{KE} + \frac{t_Z Z}{p_{ER}E}\hat{t}_Z\right]. \tag{24}$$

Again, solutions in equations (23) and (24) remain valid for the entire range of competitive settings in the energy sector ranging from monopoly ($n = 1$) to perfect competition ($n \to \infty$).

Continuing, the price of the industrial sector output is recursively given by $\hat{p}_X = \theta_E^X \hat{p}_{EX} + (1 - \theta_E^X)\hat{t}_{KX}$, and then equation (25) provides the closed-form solution as:

$$\hat{p}_X = (1 - \theta_E^X)\hat{t}_{KX} + \theta_E^X \left( \frac{n\varepsilon_{EX}}{1 + n\varepsilon_{EX}} \right) \left[ \frac{t_{EX}}{p_{EX}}\hat{t}_{EX} + \frac{p_{KE}K_E}{p_{EX}E}\hat{t}_{KE} + \frac{t_Z Z}{p_{EX}E}\hat{t}_Z \right]. \qquad (25)$$

Equation (25) demonstrates that the oligopolistic nature of the energy sector impacts output prices in other sectors via the intermediate goods market.

Next, define the relative price changes as follows: $\hat{A} \equiv \hat{p}_{ER} - \hat{p}_X$, $\hat{B} \equiv \hat{p}_{EX} - \hat{t}_{KX}$, and $\hat{C} \equiv \hat{t}_Z - \hat{t}_{KE}$. Term $\hat{A}$ represents the relative price changes of $E_R$ and $X$, where these two final products consumed by the residents and thus $\hat{A}$ must be associate with the elasticity of substitution in utility ($\sigma_U$). Similarly, term $\hat{B}$ measures relative price changes of two inputs in the industrial sector and term $\hat{C}$ provides the same for the energy sector. Therefore, in all solutions, $\hat{B}$ is associated with $\sigma_X$, and $\hat{C}$ with $\sigma_E$.

Finally, we focus on the endogenous quantity variables found in the welfare equations from Section 3; specifically, $\hat{Z}, \hat{E}_X, \hat{E}_R$, and $\hat{E}$. To start, the emission change as function of the relative input prices is given by:

$$\hat{Z} = -\sigma_U(1 - \omega_E)\varphi_R\hat{A} - \sigma_X(1 - \omega_E)(\varphi_X + \varphi_R\theta_E^X)\hat{B} - \sigma_E(1 - (1 - \omega_E)\rho_Z^E)\hat{C}. \qquad (26)$$

Thus, equation (26) has a similar structure to the specialized result in equation (21), but in the general solution all the terms on the right-hand side have ambiguous signs due to the relative price changes. Similarly, energy production (and consumption) is found to be:

$$\hat{E}_R = -\sigma_U(1 - \omega_E\varphi_R)\hat{A} - \sigma_X[\theta_E^X - \omega_E(\varphi_X + \varphi_R\theta_E^X)]\hat{B} - \sigma_E\omega_E\rho_Z^E\hat{C}, \qquad (27)$$

$$\hat{E}_X = \sigma_U\omega_E\varphi_R\hat{A} - \sigma_X[1 - \omega_E(\varphi_X + \varphi_R\theta_E^X)]\hat{B} - \sigma_E\omega_E\rho_Z^E\hat{C}. \qquad (28)$$

By inspection, $\hat{E}_R$ and $\hat{E}_X$ may have difference signs when relative price change across the two final products consumed by the residents ($\hat{A}$) is sufficiently large, and this directly impacts the price discrimination welfare effect in equation (16).

Finally, recall $\hat{E} = \varphi_X\hat{E}_X + \varphi_R\hat{E}_R$ where $\varphi_X + \varphi_R = 1$, and we find:

$$\hat{E} = -\sigma_U(1 - \omega_E)\varphi_R\hat{A} - \sigma_X(1 - \omega_E)(\varphi_X + \varphi_R\theta_E^X)\hat{B} - \sigma_E\omega_E\rho_Z^E\hat{C}. \tag{29}$$

and, by inspection, $\hat{E}$ and $\hat{Z}$ typically have the same sign (as confirmed by the numerical analysis below), as $\hat{Z} = \hat{E} - \sigma_E(1 - \rho_Z^E)\hat{C}$. The full set of solutions, including the other endogenous quantity variables, are provided in Appendix E.

# 5. Numerical Analysis

This section assigns plausible parameter values to numerically evaluate the closed-form solutions and welfare effects. The numerical analysis has four purposes. First, it helps with interpreting the analytical general equilibrium model solutions in Section 4. Second, the numerical analysis shows that the ambiguously signed price discrimination welfare effect is positive given plausible parameter values and policies such as increasing a pollution tax. Third, building the numerical analysis demonstrates how observable prices constrain the set of values that can be assigned to the unobserved parameters governing behavior of the energy sector firms. Fourth, we show how results qualitatively change if we specialize the model to perfect competition ($n \to \infty$) or monopoly ($n = 1$), and exogenously assuming these special cases leads to implausible values for unobserved parameters such as marginal cost and the residential distribution cost adder.

## 5.A Parameter Values

The parameters selected reflect the U.S. economy in 2019 as the numerical example for our model. The U.S. Energy Information Administration's State Energy Data System (SEDS) provides most of the data, including consumption and prices by production and use sectors, necessary to calculate the non-elasticity parameters (EIA, 2020). Table 1 records the relevant SEDS data as well as other parameter values for the numerical analysis. However, since the analytical model only has one energy sector then the SEDS energy sectors, such as electricity generation or petroleum production, are collapsed to a single, aggregate energy sector. The energy consumption and production aggregations are completed using energy content equivalents where energy is measured in British thermal unit (Btu), and the price aggregation is determined by the weighted average of the SEDS sector prices by total energy.[12] Given this aggregation, the energy sector produced 68.2 quadrillion Btu (QBtu) with approximately 11.4 QBtu consumed directly by the residential users at an average price of $20.8 per million Btus (mmBtu) and remainder consumed by the industrial sector at an average price of $15.2/mmBtu. All prices are in 2012 dollars. Thus, the implied total expenditure on energy consumption is nearly $1.1 trillion or 5.8 percent of 2019 U.S. GDP of $19.0 trillion (in 2012 dollars). Purchases of energy by residential users on average in the U.S. are subject to a sales tax of 8 percent while approximately half of utility purchases by industry are subject to taxation leading to a 4 percent average tax (Phillips and Ibaid, 2019).

Total capital endowment and shares of capital employed in the energy and industrial sectors are calculated from the constant return to scale production function. The energy sector employs approximately 4.3 percent of total capital while the remainder goes towards industrial production. The residential sector does not consume capital directly. Capital is the numeraire and thus a unit of capital is normalized to be equal to one dollar, but capital is still subject to taxation. However, recall that "capital" represents a composite of capital and labor, and thus we calculate a corresponding composite capital tax as follows. Cunningham *et al.* (2021) reports an average energy sector tax on actual capital of 11.5 percent while other industrial sectors are subject to an 18.2 percent tax. Goulder and Hafstead (2017) reports that the U.S. energy sector has a labor input share of 5.7 percent, and the rest of the economy has a 29.0 percent share.[13] Watson (2020) reports that on average, in 2019, labor income was subject to an 18.8 percent rate for workers with families and 29.8 percent for workers with no children, and approximately 40 percent of households have children. Using these rates and shares, we calculate $t_{KE}$=12.3 percent and $t_{KX}$=20.1 percent as the tax rates.

---

[12] Consumption Adjustments for Calculating Expenditures (Section 7) of the 2019 SEDS documentation provides technical details. SEDS adjusts its energy consumption estimates to remove process fuel and other energy consumption that has no direct fuel costs to the end-users and then summarizes overall energy expenditure and price information for each sector (EIA, 2020). To be consistent with SEDS's estimates, all energy consumption values used in this numerical analysis are after SEDS adjustments and therefore the totals reported here are smaller than actual consumption in 2019.

[13] We calculate these shares from Goulder and Hafstead (2017)'s Table 3.1 after defining the first nine sectors in that table (Oil Extraction through Petroleum Refining) as energy sectors. We assume the 2013 shares implicit in the Goulder and Hafstead (2017) data are similar enough to the 2019 values for our numerical example.

**Table 1. Data and Benchmark Parameters** (2012 Dollars)

| Symbol | Description | Value | Unit |
|--------|-------------|-------|------|
| $E_R$ | Total energy consumption by residential users | 11,428 | Trillion Btu |
| $E_X$ | Total energy consumption by industrial sector | 56,734 | Trillion Btu |
| $p_{ER}$ | Average price of energy to residents | 20.84 | \$/mmBtu |
| $p_{EX}$ | Average price of energy to industrial | 15.24 | \$/mmBtu |
| $t_{ER}$ | 8% of $p_{ER}$ | 1.67 | \$/mmBtu |
| $t_{EX}$ | 4% of $p_{EX}$ | 0.61 | \$/mmBtu |
| $\gamma$ | Marginal cost of energy (80% of $p_{EX}$) | 12.19 | \$/mmBtu |
| $\delta$ | Additional residential supply cost (11.2% of $p_{ER}$) | 2.33 | \$/mmBtu |
| $Z$ | Total $CO_2$ emissions | 5,147 | Million Metric Tons |
| $\mu$ | Marginal damage of emissions | 42.00 | \$/Metric Ton |
| $t_Z$ | Per unit emissions tax | 15.00 | \$/Metric Ton |
| $I$ | Total income as gross domestic product | 19,036 | \$Billion |
| $K_E$ | Capital employed in energy sector | 671.2 | \$Billion |
| $K_X$ | Capital employed in industrial sector | 14,932 | \$Billion |
| $t_{KE}$ | Capital tax in energy sector | 12.3 | Percent |
| $t_{KX}$ | Capital tax in industrial sector | 20.1 | Percent |
| $\omega_E$ | Share of the total capital that goes to the energy sector | 4.30 | Percent |
| $\varphi_R$ | Share of total energy products that sold to the residents | 16.77 | Percent |
| $\varphi_X$ | Share of total energy products that sold to the industry | 83.23 | Percent |
| $\varepsilon_{ER}$ | Elasticity of demand for residential energy consumption | -0.50 | |
| $\sigma_U$ | Elasticity of substitution in utility function | 0.49[b] | |
| $\sigma_E$ | Elasticity of substitution in energy production | 0.30 | |
| $n$ | Number of oligopolistic energy firms | 8.97[a] | Count |
| $\varepsilon_{EX}$ | Elasticity of demand for industrial energy consumption | -0.70[a] | |
| $\sigma_X$ | Elasticity of substitution in industrial production | 0.68[b] | |

(Notes: Capital represents a composite of all clean inputs. The capital taxes are expressed in percentages as the capital is the numeraire good. The "a" superscript denotes parameter given by the energy sector first-order conditions. The "b" superscript denotes an elasticity of substitution parameter value determined by an elasticity of demand parameter and share parameter values.)

The marginal cost of energy production varies across energy sub-sectors, and we assume it to be 80 percent of the prevailing industrial energy price ($p_{EX}$) for the benchmark case and thus equal to \$12.19/mmBtu. Sensitivity analysis on this assumption shows that the main results hold when the percentage is varied by 10 percentage points in either direction. We set the additional residential supply cost ($\delta$) at 11.2 percent to reflect the 90th percentile of residential distribution losses reported in Borenstein and Bushnell (2022) and thus equal to \$2.33/mmBtu.

Continuing, for this numerical analysis define the emissions ($Z$) as carbon dioxide ($CO_2$) emissions. In 2019, emissions totaled 5,147 million metric tons. A national $CO_2$ emission tax or price-based policy did not exist in the U.S. in 2019, but state policy in California regulated emissions across multiple sectors via a cap-and-trade regime and regional policy across a Northeastern conglomerate of states under the Regional Greenhouse Gas Initiative (RGGI) taxed carbon emissions from the electricity sector. These policies imply a cost of emissions. The model also requires an existing, initial emission tax and thus a \$15 per metric ton $CO_2$ tax is assumed (Fullerton and Karney, 2018), and leads to \$77.2 billion in implied emission costs or approximately 0.4 percent of GDP. EPA (2016) calculates the marginal damage of $CO_2$ at \$42 per metric ton and thus equation (16) shows the height of the externality Harberger triangle is \$27 per metric ton at the initial equilibrium.

Many studies provide empirical estimates for the price elasticity of resident demand for energy. Bernstein and Griffin (2006) summarize the results and indicate that the range of statistically significant estimates is from 1.10 to -1.87. For this numerical exercise, we assume $\varepsilon_{ER} = -0.50$ as the price elasticity of demand for energy by residential users. As shown in Section 2, the elasticity of substitution in utility ($\sigma_U$) is directly related to $\varepsilon_{ER}$ and the expenditure shares. We also perform sensitivity analysis on $\varepsilon_{ER}$ and find that the numerical results are similar across analyses.

Next, we assume an elasticity of substitution in energy production of $\sigma_E = 0.30$. Previous studies estimate the substitution elasticity between capital and fuel for the energy sector ranging from 0.5 to 1.2 (Zha and Ding, 2014). Continuing, Papageorgiou *et al.* (2017) finds values ranging from 2-3 for the elasticity of substitution between clean and dirty energy inputs. However, this model does not distinguish between capital and fuel as differentiated inputs in the production of energy. Rather emissions are modeled as a standalone input, and it would take much larger changes in the emission price to have the same effect as a change in the fuel price. Thus, a smaller elasticity of substitution in energy production of 0.3 is initially assumed, but sensitivity analysis shows the main result hold if the parameter value is greater than 0.10 and less than 0.60, all else equal, with most of the changes affecting the externality welfare effect.

Recall Appendix B shows the relationship between the energy prices and associated elasticities is governed by following first-order conditions: $n(p_{EX} - t_{EX}) + \frac{p_{EX}}{\varepsilon_{EX}} = n\gamma$ and $n(p_{ER} - \delta - t_{ER}) + \frac{p_{ER}}{\varepsilon_{ER}} = n\gamma$. Given assumptions regarding marginal cost and price elasticity of resident demand for energy, then the two equations solve for the implied number of firms ($n$) and price elasticity of demand for industrial energy use $(\varepsilon_{EX})$, and found to be 8.97 and -0.70, respectively. A value of $\varepsilon_{EX} = -0.70$ fits in the range of estimates in literature (Bernstein and Griffin, 2006) and thus is a plausible value. The number of firms is nearly 9 meaning the data implies that energy market consists of firms that behave equivalently to 9 identical firms conditional on the model structure. Again, the number of firms can be interpreted as a competition index in this model. Sensitivity analysis below varies the assumption regarding marginal cost and reports the impacts on $n$ and $\varepsilon_{EX}$. Alternatively, the numerical analysis could set $\varepsilon_{EX}$ exogenously and use first-order conditions above to find the number of firms and marginal cost, but numerical results would be identical holding all else equal. Again, as shown in Section 2, the elasticity of substitution in industrial production $(\sigma_X)$ is directly related to $\varepsilon_{EX}$ and the expenditure shares.

## 5.B Numerical Results

### 5.B.1 Emission Tax, Revenue Recycling, and Two-Part Instrument

This model with oligopolistic energy producers has three (potential) distortions in general equilibrium – insufficient production due to market power, negative externality when the prevailing tax is lower than the marginal damage, and price discrimination across user types. In this multiple distortion setting, using a single instrument may help correct one distortion while exacerbating another distortion. For example, a higher emission tax reduces the negative externality, but the resulting higher product price can further exacerbate the production distortion. However, revenue recycling from emission tax revenue can be used to help correct distortions (Goulder, 1995; Bovenberg and Goulder, 1996; Goulder *et al.*, 1997; Fullerton and Metcalf, 2001). The presence of three distortions and many taxes in this model provides a multitude of options when implementing revenue recycling.

The benchmark case (Case 1.0) increases the emission tax by 10 percent and Table 2 reports the results of this policy change. Panel A records the policy change with the non-zero, exogenous policy highlighted in bold $(\hat{t}_Z = 10\% > 0)$. The other taxes are fixed with respect to the numeraire and thus recorded as zeros in the table. Then, panel B records the resulting price and quantity changes due to the policy change; for example, the price of energy for industrial users increases more $(\hat{p}_{EX} = 0.88\%)$ than the price for residential customers $(\hat{p}_{ER} = 0.70\%)$, and thus shrinks the relative gap in energy prices. Also, emissions fall by 3.27 percent because of the 10 percent emission tax increase. Finally, panel C reports the welfare changes using the all-tax version of equation (16). Overall, the benchmark case improves welfare by \$3.3 billion (in 2012 dollars). Most welfare gains come from the reduced negative externality although the price discrimination effect is also reduced – and thus welfare increases – as the gap between energy prices shrinks. However, the emission tax increase leads to greater energy sector production costs and thus further reduces equilibrium energy output yielding a larger output distortion with a negative welfare effect of \$1.2 billion.[14] The Case 1.0 benchmark provides an example of Theorem 1 $\left(\hat{E}_X < \hat{E}_R < 0\right)$ yielding a positive price discrimination effect and a negative oligopoly output effect on welfare. However, Case 1.0 fails to satisfy the necessary condition of Theorem 2 $\left(\hat{E}_R > 0\right)$ and thus oligopoly output effect is larger in absolute value than the price discrimination effect. Energy firm profits increase from an emission tax in this benchmark case.[15]

Next, Cases 2.0-4.0 investigate the effects of revenue recycling and a two-part instrument. To maintain comparability, Cases 2.0-4.0 maintain a balanced budget constraint $(\hat{T} = 0)$.

In Case 2.0, the emission tax revenue is used to reduce $t_{ER}$ by 28.33 percent and as a result the residential energy price falls by 2.22 percent. This fall in the residential price means the price discrimination distortion greatly diminishes and raises welfare by over \$900 million. The necessary condition of Theorem 2 is satisfied $\left(\hat{E}_R > 0\right)$ and the sufficient condition is also satisfied as the price discrimination welfare gain more than fully counteracts the negative oligopoly output effect. However, total energy use is higher than in the benchmark case and thus yields a smaller gain from externality correction.

For Case 3.0, total emission reductions are fixed at the benchmark case with $\hat{Z} = -3.27$ percent but tax revenue is still used to reduce $t_{ER}$. Here, the residential energy price falls by 2.38 percent. As in Case 2.0, the price discrimination welfare gain more than fully counteracts the negative oligopoly output effect, but the total welfare gain is larger due to a larger externality reduction from the higher emission tax ($\hat{\tau}_Z = 10.76$). In both revenue recycling cases, energy firm profits fall.

---

[14] Appendix D shows welfare changes with disaggregated capital tax effects. In general, the tax effects are small in magnitude relative to the main price effects.

[15] The initial level of energy profit is unobserved and in the numerical analysis subject to model parameterization via the marginal cost. Therefore, the percentage change in profits cannot be compared across all cases. However, the sign of the profit change is determined regardless of level and thus the results highlight the profit sign change only.

Then, Case 4.0 explores a two-part instrument where the revenue from raising the tax on $E_X$ is used to cut tax on $K_E$ and thus changes the relative price in the energy sector.[16] The marginal cost of energy production falls by $\hat{\gamma} = -8.97$ percent as the capital tax in the energy sector falls. Again, the necessary and sufficient conditions for Theorem 2 hold. The theoretical importance of Case 4.0 is that it mimics the classic two-part instrument for correcting an externality. An emission tax typically generates abatement via two levers: an output effect and a substitution effect. The substitution effect occurs as firms switch to cleaner inputs as the relative price of emissions increase via the tax. The equilibrium price of the "dirty" good must increase too and thus the output effect arises as fewer units of the dirty good are produced and sold. That is, an emission tax lowers emissions per unit of output and the amount of output sold, and thus a two-part instrument mimics these two effects by taxing output and subsidizing the clean input. Case 4.0 also finds a decrease in energy firm profits. The numerical analysis implies another interesting possibility that, under current model parameterization, the two-part instrument Case 4.0 without increasing the emission tax directly even generates a higher welfare gain than the other cases.

---

[16] Recall, while most changes in taxes, such as $\hat{t}_{ER}$ and $\hat{t}_{EX}$, are defined relative to their initial tax, the capital tax changes, such as $\hat{t}_{KE}$, are defined relative to the price of the capital.

**Table 2. Emission Tax, Revenue Recycling, and Two-Part Instrument**

| Case: | 1.0 | 2.0 | 3.0 | 4.0 |
|---|---|---|---|---|
| **Panel A: Exogenous Policy Change (in percentage)** | | | | |
| $\hat{t}_Z$ | **10.00** | **10.00** | 10.76 | 0.00 |
| $\hat{t}_{ER}$ | 0.00 | -28.33 | -30.46 | 0.00 |
| $\hat{t}_{EX}$ | 0.00 | 0.00 | 0.00 | 221.91 |
| $\hat{t}_{KE}$ | 0.00 | 0.00 | 0.00 | -9.89 |
| **Panel B: Closed-Form Solutions (in percent)** | | | | |
| $\hat{\gamma}$ | 0.93 | 0.93 | 1.00 | -8.97 |
| $\hat{p}_{EX}$ | 0.88 | 0.88 | 0.95 | 2.02 |
| $\hat{p}_{ER}$ | 0.70 | -2.22 | -2.38 | -6.75 |
| $\hat{p}_X$ | 0.04 | 0.04 | 0.04 | 0.09 |
| $\hat{K}_X$ | 0.01 | 0.00 | 0.00 | 0.01 |
| $\hat{K}_E$ | -0.27 | -0.04 | -0.04 | -0.30 |
| $\hat{E}$ | -0.55 | -0.32 | -0.34 | -0.58 |
| $\hat{E}_R$ | -0.34 | 1.09 | 1.17 | 3.33 |
| $\hat{E}_X$ | -0.59 | -0.60 | -0.65 | -1.37 |
| $\hat{X}$ | -0.02 | -0.03 | -0.03 | -0.05 |
| $\hat{Z}$ | -3.27 | -3.04 | **-3.27** | **-3.27** |
| $\hat{\Pi}_E$ | 0.31 | -0.11 | -0.11 | -0.47 |
| $\hat{T}$ | 0.16 | **0.00** | **0.00** | **0.00** |
| **Panel C: Welfare Effects (in 2012$ million)** | | | | |
| Total: | 3327.5 | 4280.1 | 4604.5 | 5631.1 |
| | | | | |
| Oligopoly Output: | -1175.8 | -681.4 | -733.5 | -1240.5 |
| Price Discrimination: | 133.5 | 900.3 | 968.2 | 2501.8 |
| Externality: | 4369.8 | 4061.2 | **4369.8** | **4369.8** |

(Notes: The bold cells indicate the exogenous policy parameters in panels A and B. In Panel C, cells are independent rounded such that sum totals may not equal the constituent parts. See Table D1 in Appendix D for the differentiated capital tax welfare effects.)

## 5.B.2 Parameter Sensitivity

While many of the parameters and underlying data presented in Table 1 are well known, some other parameters are subject to greater uncertainty; specifically, the marginal cost of energy production ($\gamma$), the elasticity of substitution in the production of energy ($\sigma_E$), and residential price elasticity of demand for energy ($\varepsilon_{ER}$). Table 3 provides results for sensitivity analysis on these parameters. The first column of Table 3 repeats the benchmark Case 1.0 results (i.e., the "base" case). Then, Cases 1.1-1.6 perform the sensitivity checks on key parameters while maintaining a 10 percent emission tax increase. Cases 1.1 and 1.2 vary the initial marginal cost parameter to $10.67 or 70 percent of $p_{EX}$ ("Low $\gamma$"), and $13.71 or 90 percent of $p_{EX}$ ("Hight $\gamma$"), respectively. Recall the benchmark marginal energy cost is set at $12.19 or 80 percent of the prevailing industrial energy price. For Cases 1.3 and 1.4, the values for $\sigma_E$ are set at 0.1 and 0.6, respectively, compared to the benchmark value of 0.3 for the substitution elasticity. For Cases 1.5 and 1.6, the values of $\varepsilon_{ER}$ are varied. Recall that $n$ and $\varepsilon_{ER}$ are solved for after other parameters are set and thus change accordingly across cases.

Importantly, changing parameter values changes the initial equilibrium and thus sensitivity cases cannot be directly compared to other sensitivity cases. Table 3 demonstrates this point by showing the initial number of equivalent oligopolistic firms. For Cases 1.3 and 1.4, however, changing the energy sector substitution elasticity does not affect the other elasticities or number of firms since $\sigma_E$ does not appear in the energy sectors first-order conditions.

The results for Cases 1.1 and 1.2 are similar to the benchmark results except that profits increase more in the low marginal cost case. The Case 1.1 initial equilibrium now has inelastic industrial demand and fewer firms relative to the benchmark case, and both could contribute to the larger profit increase.

Cases 1.3 and 1.4 have different welfare effects with a small elasticity of substitution in the energy sector leading to small welfare gains while a higher $\sigma_E$ leads to large welfare gains. A small $\sigma_E$ limits the energy sector ability to reduce per unit emissions via a substitution effect and thus relies almost exclusively on the output effect to lower emissions. Thus, for a given tax increase a small elasticity of substitution yields less emission reductions. In contrast, the higher $\sigma_E$ in Case 1.4 means the firms can more easily abatement via substitution and thus a given tax increase is more effective at reduction emissions and increasing welfare, all else equal. Interestingly, both cases still find an increase in energy firm profits.

Cases 1.5 and 1.6 show that varying the residential demand elasticity has large impact on the number of firms in the model but the overall welfare change remain largely unaffected.

**Table 3. Parameter Sensitivity ($\hat{t}_Z = 10$; $\hat{t}_{ER} = \hat{t}_{EX} = \hat{t}_{KE} = \hat{t}_{KX} = 0$)**

| Case: | 1.0 | 1.1 | 1.2 | 1.3 | 1.4 | 1.5 | 1.6 |
|---|---|---|---|---|---|---|---|
| Description: | Base | Low $\gamma$ | High $\gamma$ | Low $\sigma_E$ | High $\sigma_E$ | Low $\varepsilon_{ER}$ | High $\varepsilon_{ER}$ |
| **Panel A: New Parameter Values** | | | | | | | |
| $\gamma$ | 12.19 | **10.67** | **13.71** | 12.19 | 12.19 | 12.19 | 12.19 |
| $\sigma_E$ | 0.30 | 0.30 | 0.30 | **0.10** | **0.60** | 0.30 | 0.30 |
| $n$ | 8.97 | 6.75 | 13.35 | 8.97 | 8.97 | 17.94 | 5.98 |
| $\varepsilon_{ER}$ | -0.50 | -0.50 | -0.50 | -0.50 | -0.50 | **-0.25** | **-0.75** |
| $\varepsilon_{EX}$ | -0.70 | -0.57 | -1.25 | -0.70 | -0.70 | -0.35 | -1.05 |
| **Panel B: Closed-Form Solutions (in percent)** | | | | | | | |
| $\hat{\gamma}$ | 0.93 | 1.06 | 0.83 | 0.93 | 0.93 | 0.93 | 0.93 |
| $\hat{p}_{EX}$ | 0.88 | 1.00 | 0.79 | 0.88 | 0.88 | 0.88 | 0.88 |
| $\hat{p}_{ER}$ | 0.70 | 0.77 | 0.64 | 0.70 | 0.70 | 0.70 | 0.70 |
| $\hat{p}_X$ | 0.04 | 0.05 | 0.04 | 0.04 | 0.04 | 0.04 | 0.04 |
| $\hat{K}_X$ | 0.01 | 0.01 | 0.03 | 0.02 | 0.00 | 0.00 | 0.03 |
| $\hat{K}_E$ | -0.27 | -0.20 | -0.61 | -0.45 | -0.00 | 0.02 | -0.56 |
| $\hat{E}$ | -0.55 | -0.52 | -0.86 | -0.54 | -0.56 | -0.26 | -0.84 |
| $\hat{E}_R$ | -0.34 | -0.38 | -0.31 | -0.33 | -0.35 | -0.17 | -0.51 |
| $\hat{E}_X$ | -0.59 | -0.54 | -0.97 | -0.58 | -0.60 | -0.28 | -0.90 |
| $\hat{X}$ | -0.02 | -0.02 | -0.01 | -0.01 | -0.03 | -0.01 | -0.02 |
| $\hat{Z}$ | -3.27 | -3.20 | -3.61 | -1.45 | -6.00 | -2.98 | -3.56 |
| $\hat{\Pi}_E$ | 0.31 | 0.45 | 0.03 | 0.32 | 0.30 | 0.58 | 0.04 |
| $\hat{T}$ | 0.16 | 0.16 | 0.15 | 0.20 | 0.09 | 0.16 | 0.15 |
| **Panel C: Welfare Effects (in 2012\$ million)** | | | | | | | |
| Total: | 3327.5 | 2684.8 | 4287.7 | 910.1 | 6953.6 | 3482.5 | 3172.5 |
| Oligopoly Output: | -1175.8 | -1675.5 | -881.2 | -1158.7 | -1201.5 | -562.6 | -1789.0 |
| Price Discrimination: | 133.5 | 89.1 | 347.9 | 133.5 | 133.5 | 58.1 | 208.8 |
| Externality: | 4369.8 | 4271.2 | 4821.0 | 1935.3 | 8021.6 | 3987.0 | 4752.7 |

(Notes: In panel A, the bold cells indicate the new parameter assumptions relative to the base case. For Cases 1.1 and 1.2, the initial values for $\gamma$ are set at 10.67 (70 percent) and 13.72 (90 percent), respectively, compared to the benchmark value of 12.20 (80 percent of industrial sector energy price). Then, for Cases 1.3 and 1.4, the values for $\sigma_E$ are set at 0.1 and 0.6, respectively, compared to the benchmark value of 0.3 for the substitution elasticity. In Panel C, cells are independent rounded such that sum totals may not equal the constituent parts. See Table D2 in Appendix D for the differentiated capital tax welfare effects.)

## 5.B.3 Structural Sensitivity

Table 4 provides a structural sensitivity analysis by forcing a monopolistic energy sector (Case 1.7) and perfect competition energy sector (Case 1.8). That is, for Case 1.7, we set $n = 1$ and for Case 1.8, $n \to \infty$, but maintaining policy of a 10 percent emission tax increase. The first order conditions must still hold and thus elasticity values or the marginal cost assumption must be adjusted. The monopoly case then requires larger elasticity values of -4.49 and -6.25 for $\varepsilon_{ER}$ and $\varepsilon_{EX}$, respectively. Thus, Case 1.7 shows a very large decrease in emissions, but that welfare gain is more than entirely offset by an increase in the output distortion. Perfect competition requires $(p_{EX} - t_{EX}) = (p_{ER} - \delta - t_{EX}) = \gamma$ due to the zero-profits condition. (Note: $\widehat{\Pi}_E$ is undefined as the denominator is zero.) Then, Case 1.8 finds only a small price discrimination effect due to the prevailing tax distortions meaning nearly all the overall welfare gain comes from the externality correction. Also, despite the perfect competition setting, the market has a non-zero initial output distortion due to prevailing tax distortions.

From Table 2, recall that the two-part instrument policy scenario, Case 4.0, yields the largest welfare gain. Table 4 then provides structural sensitivity for that policy scenario too. Thus, Cases 4.7 and 4.8 build on Case 4.0 from Table 2, but exogenously impose monopoly and perfect competition market structures, respectively. To keep cases comparable, the tax changes from Case 4.0 are kept for Cases 4.7 and 4.8 with the tax on capital in the energy sector falling while the tax on energy sector sales to the industrial sector rising. Table 4 shows qualitatively large difference between Cases 1.7 and 4.7 as the two-part instrument would significantly raise total welfare as the price discrimination distortion in the latter case yield substantial welfare gains, while the oligopoly output and externality welfare effects are largely unchanged. In contrast, Cases 1.8 and 4.8 have similar overall welfare gains although the latter case has a smaller externality correction.

Overall, Table 4 demonstrates that the market structure matter for environmental policy analysis. That is, Case 1.0 is qualitatively different from Cases 1.7 and 1.8; similarly, Case 4.0 is qualitatively different from Cases 4.7 and 4.8. Thus, models that assume either perfect competition or monopoly provide answers that are not only different in size but also distributional impacts. This model has one representative consumer, but it is likely that the different distortions affect different groups differently. For example, price discrimination effects increase energy prices to residential consumers and that impacts lower income deciles that spend a higher fraction of their income on goods like electricity compared to the higher income deciles (Cronin *et al.*, 2019). Also, if environmental amenities are normal (or even luxury) goods, then higher income individuals benefit more from the externality correction.

**Table 4. Structural Sensitivity**

| Case: | 1.0 | 1.7 | 1.8 | 4.0 | 4.7 | 4.8 |
|---|---|---|---|---|---|---|
| Description: | Base Tax | Monopoly | Perfect Comp. | Base Two-Part | Monopoly | Perfect Comp. |
| **Panel A: Exogenous Policy Change (in percentage)** | | | | | | |
| $\hat{t}_Z$ | **10.00** | **10.00** | **10.00** | 0.00 | 0.00 | 0.00 |
| $\hat{t}_{EX}$ | 0.00 | 0.00 | 0.00 | **221.91** | **221.91** | **221.91** |
| $\hat{t}_{KE}$ | 0.00 | 0.00 | 0.00 | **-9.89** | **-9.89** | **-9.89** |
| **Panel B: New Parameter Values** | | | | | | |
| $\gamma$ | 12.19 | 12.19 | 14.63 | 12.19 | 12.19 | 14.63 |
| $\sigma_E$ | 0.30 | 0.30 | 0.30 | 0.30 | 0.30 | 0.30 |
| $n$ | 8.97 | **1** | $\infty$ | 8.97 | **1** | $\infty$ |
| $\varepsilon_{ER}$ | -0.50 | -4.49 | -0.50 | -0.50 | -4.49 | -0.50 |
| $\varepsilon_{EX}$ | -0.70 | -6.25 | -0.70 | -0.70 | -6.25 | -0.70 |
| **Panel C: Closed-Form Solutions (in percent)** | | | | | | |
| $\hat{y}$ | 0.93 | 0.93 | 0.77 | -8.97 | -8.97 | -9.12 |
| $\hat{p}_{EX}$ | 0.88 | 0.88 | 0.74 | 2.02 | 2.02 | 0.12 |
| $\hat{p}_{ER}$ | 0.70 | 0.70 | 0.54 | -6.75 | -6.75 | -6.40 |
| $\hat{p}_X$ | 0.04 | 0.04 | 0.03 | 0.09 | 0.09 | 0.01 |
| $\hat{K}_X$ | 0.01 | 0.22 | 0.01 | 0.01 | 0.24 | -0.04 |
| $\hat{K}_E$ | -0.27 | -4.84 | -0.22 | -0.30 | -5.34 | 0.66 |
| $\hat{E}$ | -0.55 | -5.12 | -0.45 | -0.58 | 5.26 | 0.43 |
| $\hat{E}_R$ | -0.34 | -3.03 | -0.26 | 3.33 | 30.65 | 3.12 |
| $\hat{E}_X$ | -0.59 | -5.54 | -0.49 | -1.37 | -12.96 | -0.12 |
| $\hat{X}$ | -0.02 | -0.05 | -0.01 | -0.05 | -0.37 | -0.04 |
| $\hat{Z}$ | -3.27 | -7.84 | -3.22 | **-3.27** | -8.31 | -2.31 |
| $\widehat{\Pi}_E$ | 0.31 | -4.01 | n/a | -0.47 | -1.25 | n/a |
| $\hat{T}$ | 0.16 | 0.05 | 0.16 | **0.00** | 0.00 | -0.49 |
| **Panel D: Welfare Effects (in 2012$ million)** | | | | | | |
| Total: | 3327.5 | 856.1 | 4265.2 | 5631.1 | 22282.1 | 4967.8 |
| Oligopoly Output: | -1175.8 | -10950.5 | -165.4 | -1240.5 | -12026.4 | 154.8 |
| Price Discrimination: | 133.5 | 1334.5 | 123.7 | 2501.8 | 23205.1 | 1726.4 |
| Externality: | 4369.8 | 10472.1 | 4306.9 | **4369.8** | 11103.4 | 3086.5 |

(Notes: The bold cells in panel A indicate the exogenous policy parameters. In panel B, the bold cells indicate the new parameter assumptions relative to the base case. In Panel C, cells are independent rounded such that sum totals may not equal the constituent parts. See Table D3 in Appendix D for the differentiated capital tax welfare effects. For the perfect competition cases, the change in profits is not defined as profits are always zero and thus an "n/a" entry in the table.)

# 6. Conclusion

This study constructs a novel analytical general equilibrium model with an oligopolistic energy sector to investigate environmental policy under market power with price discrimination. The oligopolistic firms produce energy sold to residential consumers as a final good and sold to industrial firms as an intermediate good. However, the homogenous energy is differentially priced across users via third-degree price discrimination. We find closed-form solutions for the effects of environmental policy on prices and quantities. In addition, we derive a welfare equation that disaggregates effects across the three prevailing distortions: oligopoly output, price discrimination, and externality. In this setting, numerical analysis demonstrates that otherwise equivalent environmental policies – such as an emission tax and a two-part instrument – have different impacts on the endogenous variables and welfare outcomes. Also, the oligopoly output and price discrimination effects can have offsetting welfare effects leaving the externality correction as the net effect of environmental policy. These offsetting effects ameliorate the concern that environmental policy in an imperfect market with multiple distortions reduces the net welfare gain from the externality correction.

Appendices

## Appendix A: Parameter Definitions

$\gamma$, marginal cost of energy sector production.

$\delta$, additional distribution cost of selling energy goods to residential users.

$\omega_E \equiv K_E / \overline{K}$, share of total capital employed in the energy sector.

$\theta_E^X \equiv \frac{p_{EX} E_X}{p_X X}$, share of total revenue that industrial firms spend on energy inputs.

$\theta_K^X \equiv \frac{p_{KX} K_X}{p_X X}$, share of total revenue that industrial firms spend on capital inputs.

$\theta_E^R \equiv p_{ER} E_R / I$, share of total income that residential consumers spend on energy.

$\rho_Z^E \equiv \frac{t_Z Z}{\gamma E}$, ratio of total emission taxes to total marginal energy production costs.

$\rho_K^E \equiv \frac{p_{KE} K_E}{\gamma E}$, ratio of total capital costs to total marginal energy production costs.

$\sigma_U \equiv \frac{d ln(X/E_R)}{d ln(p_{ER}/p_X)}$, elasticity of substitution between final goods in utility function.

$\sigma_X \equiv \frac{d ln(E_X/K_X)}{d ln(p_{KX}/p_{EX})}$, elasticity of substitution in production of the industrial good.

$\sigma_E \equiv \frac{d ln(Z/K_E)}{d ln(p_{KE}/t_Z)}$, elasticity of substitution in production of energy.

$\varepsilon_{ER} \equiv \frac{\partial E_R / E_R}{\partial p_{ER} / p_{ER}}$, residential consumers' price elasticity of demand for energy.

$\varepsilon_{EX} \equiv \frac{\partial E_X / E_X}{\partial p_{EX} / p_{EX}}$, industrial sector's price elasticity of demand for the energy.

$\varphi_X \equiv E_X / E$, share of total energy products that sold to industrial firms as intermediate goods.

$\varphi_R \equiv E_R / E$, share of total energy products that sold to the residents as final goods.

$\beta_E \equiv q_K / p_{KE}$, non-tax share of the total capital price in the energy sector.

$\beta_X \equiv q_K / p_{KX}$, non-tax share of the total capital price in the industrial sector.

## Appendix B: Oligopoly Energy Firms' Profit Maximization

Let the production function for firm $i \in (1, \ldots, n)$ in the oligopolistic energy sector be given as $E_i = E_i(K_{Ei}, Z_i)$. Assume firms engage in quantity competition via the Cournot model. Define the inverse demand function of the residential consumers as $p_{ER} = f_R(E_R)$, where $E_R = \sum_{i=1}^{n} E_{Ri}$ such that $E_{Ri}$ is the amount sold by firm $i$ to the residents. Similarly, define the industrial sector's inverse demand function as $p_{EX} = f_X(E_X)$, where $E_X = \sum_{i=1}^{n} E_{Xi}$, and note that $E_i = E_{Xi} + E_{Ri}$. Then, firm $i$ has the profit maximization problem:

$$\max_{E_{Xi}, E_{Ri}, K_{Ei}, Z_i} [f_X(E_{Xi} + E_{-Xi}) - t_{EX}]E_{Xi} + [f_R(E_{Ri} + E_{-Ri}) - \delta_i - t_{ER}]E_{Ri} - p_{KE}K_{Ei} - t_Z Z_i$$

$$s.t. \ E_{Xi} + E_{Ri} = E_i(K_{Ei}, Z_i),$$

where $E_{-Xi}$ and $E_{-Ri}$ represent the total production from the other firms. The optimization's first-order conditions (FOCs) are found to be:

$$[K_E]: p_{KE} = \gamma_i \frac{\partial E_i}{\partial K_{Ei}},$$

$$[Z]: t_Z = \gamma_i \frac{\partial E_i}{\partial Z_i},$$

$$[E_{Xi}]: \overbrace{f_X(E_{Xi} + E_{-Xi}) - t_{EX} + \frac{\partial f_X(E_{Xi} + E_{-Xi})}{\partial E_{Xi}} E_{Xi}}^{MR} = \overbrace{\gamma_i}^{MC},$$

$$[E_{Ri}]: \overbrace{f_R(E_{Ri} + E_{-Ri}) - \delta_i - t_{ER} + \frac{\partial f_R(E_{Ri} + E_{-Ri})}{\partial E_{Ri}} E_{Ri}}^{MR} = \overbrace{\gamma_i}^{MC},$$

where $\gamma_i$ is the Lagrange multiplier for firm $i$ and can be interpreted as the marginal cost of producing one more unit of output. Observe the $[E_{Xi}]$ and $[E_{Ri}]$ FOCs equate the marginal cost (MC) and marginal revenue (MR) of producing energy and then selling it to either the industrial sector or residential consumers, where $\delta_i$ is the additional distribution cost of selling to residential users.

Let the Cournot output for each of the $n$-indentical firms in oligopolistic energy sector be given $E_{Xi}^*$ and $E_{Ri}^*$, and let $\gamma_i$ be the marginal cost of firm $i$. Summing over the $n$ equations for the FOC $[E_{Ri}]$ yields:

$$n\varepsilon_{ER}(p_{ER}^* - \delta_i - t_{ER}) + p_{ER}^* = \varepsilon_{ER} \sum_{i=1}^{n} \gamma_i,$$

where $p_{ER}^* = f_R(E_R^*)$ is the equilibrium price of energy products as an intermediate good and define $\varepsilon_{ER} \equiv (\partial E_R/E_R)/(\partial p_{ER}/p_{ER})$. Since all firms in the energy sector are identical, then $\gamma_i = \gamma$ and $\delta_i = \delta$. After substituting, it follows that $n\varepsilon_{ER}(p_{ER}^* - \delta - t_{ER}) + p_{ER}^* = n\varepsilon_{ER}\gamma$ and we rewrite to find $0 \leq (p_{ER}^* - \delta - t_{ER} - \gamma)/p_{ER}^* = -1/n\varepsilon_{ER} < 1$, which implies $\varepsilon_{ER} < -1/n$.

We next drop the "*" notation since $E_{Ri}$ and $p_{ER}$ come from the equilibrium conditions, and rearrange to find:

$$p_{ER} = \left(\frac{n\varepsilon_{ER}}{1 + n\varepsilon_{ER}}\right)(\gamma + \delta + t_{ER}),$$

and next we totally differentiate the above equation. Define $\hat{t}_{ER} = dt_{ER}/t_{ER}$ as the proportional change of per unit tax on intermediate goods and $\hat{\gamma}$ and the proportional change in marginal costs, but the additional residential distribution cost parameter is fixed so $d\delta = 0$ and thus drops out of the log-linear equations. Substituting and rearranging leads to equation (9). We repeat this process and make analogous definitions for the industrial sector to yield equation (8).

Continuing, due to symmetry, the other inputs and outputs for the energy firms must be identical too: $\hat{E}_i = \hat{E}_j$ and $\hat{Z}_i = \hat{Z}_j$ for all firms $i, j \in (1, \ldots, n)$. Therefore, $\hat{E} = \sum_{i=1}^{n} \frac{E_i}{E} \hat{E}_i = \sum_{i=1}^{n} \frac{1}{n} \hat{E}_i = \hat{E}_i$, and $\hat{K}_E = \hat{K}_{Ei}$, $\hat{Z} = \hat{Z}_i$, $\hat{E}_X = \hat{E}_{Xi}$ and $\hat{E}_R = \hat{E}_{Ri}$; that is, knowing the proportional changes for a given firm in the energy sector reveals the proportional changes for the entire sector. Then, totally differentiating firm $i$'s production function, $E_i = E_i(K_{Ei}, Z_i)$, and substituting the remaining FOCs and switching the firm-to-sector proportional changes leads to equation (10). CRS production implies $\gamma_i E_i = p_{KE} K_{Ei} + t_Z Z_i$, and then totally differentiating provides and again switching the firm-to-sector proportional changes yields equation (11).

Marginal cost ($\gamma$) in this model is the long-run, full cost of quantity adjustment. Although the energy sector, including its distribution utilities, was a classic example of a natural monopoly (Posner, 1969; Baumol, 1977), the average size of generators, and the fixed cost of building a single power generator, becomes much smaller due to the development of various technologies, like renewable energy and distributed generation (Pepermans *et al.*, 2005; Borenstein and Bushnell, 2015;). Christensen and Greene (1976) also find that firms in the U.S. electricity generation are operating in the flat area of the average cost curve by 1970 and conclude that a few huge firms are not necessary for efficient production, though there were significant scale economies available to the industry. As the consumers' demand from energy sector has increased, to supply an economy like a state or nation needs many power plants. The aggregation of numbers of generator units has regularizing effects and thus can make the average production set almost convex (Mas-Colell *et al.*, 1995). Moreover, the log-linearized general equilibrium model captures the relatively long-term effects from one equilibria to the next with small enough tax or other exogenous changes. In our case, the whole energy sector in a large economy like the U.S. will need to retire or add many generators or facilities when facing policy shocks. The marginal cost of energy products in this model is not only the marginal cost of producing one more unit of electricity or gasoline at given production capacity. Rather, the long-run marginal cost defined in this model covers both incremental capacity changes from new or retiring plants, and quantity shifts after implement of new environmental policies. Since the energy sector has many units then one can assume a sector-wide aggregate production function exhibits constant returns to scale.

## Appendix C: Energy Firm Profits and Lump-Sum Transfer

The total net profit obtained by the oligopolistic energy producers can be written as $\Pi_E = (p_{ER} - \delta - t_{ER} - \gamma)E_R + (p_{EX} - t_{EX} - \gamma)E_X$. Totally differentiating the profit function yields:

$$\widehat{\Pi}_E = \frac{E_R}{\Pi_E}\big[(p_{ER} - \delta - t_{ER} - \gamma)\hat{E}_R + p_{ER}\hat{p}_{ER} - t_{ER}\hat{t}_{ER} - \gamma\hat{\gamma}\big]$$

$$+ \frac{E_X}{\Pi_E}\big[(p_{EX} - t_{EX} - \gamma)\hat{E}_X + p_{EX}\hat{p}_{EX} - t_{EX}\hat{t}_{EX} - \gamma\hat{\gamma}\big]$$

recalling the budget constraint is given by $I \equiv q_K\overline{K} + \Pi_E + \delta E_R + T = p_X X + p_{ER}E_R$. Positive profits wedges of $(p_{ER} - \delta - t_{ER} - \gamma) > 0$ and $(p_{EX} - t_{EX} - \gamma) > 0$ means profit changes are correlated with output, all else equal, but the other terms change the size of the profit wedge per unit sold. Here, $\delta$ appears in $\widehat{\Pi}_E$ since the residential energy distribution adder affects the per unit profit margin for residential energy sales.

The lump-sum transfer must be equal net tax revenue collected by the government, that is, $T = t_{EX}E_X + t_{ER}E_R + t_{KE}K_E + t_{KX}K_X + t_Z Z$. Totally differentiate this condition obtains:

$$\hat{T} = \theta_{EX}^T(\hat{t}_{EX} + \hat{E}_X) + \theta_{ER}^T(\hat{t}_{ER} + \hat{E}_R) + \theta_{KE}^T\left(\hat{K}_E + \frac{p_{KE}}{t_{KE}}\hat{t}_{KE}\right) + \theta_{KX}^T\left(\hat{K}_X + \frac{p_{KX}}{t_{KX}}\hat{t}_{KX}\right) +$$

$$\theta_Z^T(\hat{t}_Z + \hat{Z}),$$

where $\theta_{EX}^T \equiv t_{EX}E_X/T$ and $\theta_{ER}^T \equiv t_{ER}E_R/T$, $\theta_{KE}^T \equiv t_{KE}K_E/T$ and $\theta_{KX}^T \equiv t_{KX}K_X/T$, and $\theta_Z^T \equiv t_Z Z/T$ are the shares of total taxes from energy sales, capital taxes, and the emission tax, respectively.

## Appendix D: Derivation of Welfare Equations

The residents' problem is to maximize their utility:

$$\max_{X,E_R} U(X, E_R; Z) \text{ s.t. } I = p_X X + p_{ER}E_R.$$

Note that atomistic residents cannot choose the level of externality ($Z$). The first-order conditions to the above optimization problem are given:

$$[X]: \frac{\partial U}{\partial X} = \lambda p_X > 0,$$

$$[E_R]: \frac{\partial U}{\partial E_R} = \lambda p_{ER} > 0$$

where $\lambda > 0$ is the Lagrange multiplier on the budget constraint. To start, totally differentiate the representative consumer's utility function $U(X, E_R; Z)$ and substitute in the FOCs from the utility maximization problem to find:

$$dU = \frac{\partial U}{\partial X} dX + \frac{\partial U}{\partial E_R} dE_R + \frac{\partial U}{\partial Z} dZ = \lambda p_X dX + \lambda p_{ER} dE_R + \frac{\partial U}{\partial Z} dZ \qquad (D1)$$

where $\lambda$ is the shadow price on the budget constraint. Next, totally differentiate the production function $X = X(K_X, E_X)$ and substitute in the FOCs from the profit maximization problem to get:

$$dX = \frac{\partial X}{\partial K_X} dK_X + \frac{\partial X}{\partial E_X} dE_X = \frac{p_{KX}}{p_X} dK_X + \frac{p_{EX}}{p_X} dE_X. \qquad (D2)$$

Continuing, plug equation (D2) into equation (D1) to yield:

$$dU = \lambda (p_{KX} dK_X + p_{EX} dE_X) + \lambda p_{ER} dE_R + \frac{\partial U}{\partial Z} dZ.$$

Dividing both sides by $\lambda I$ gives:

$$\frac{dU}{\lambda I} = \frac{1}{I} (p_{KX} dK_X + p_{EX} dE_X) + \frac{p_{ER}}{I} dE_R - \mu \frac{Z}{I} \hat{Z}. \qquad (D3)$$

Since $E_R$ and $E_X$ are identical goods from the same production function $E = E(K_E, Z)$, it must be that $E = E_X + E_R$. Also, recall the capital resource constraint $K_E + K_X = \overline{K}$. Thus, rewrite the above production function of energy as:

$$E_X + E_R = E(\overline{K} - K_X, Z).$$

Then, totally differentiating the above equation and substitute in the FOCs to find:

$$dE_X + dE_R = -\frac{\partial E}{\partial K_E} dK_X + \frac{\partial E}{\partial Z} dZ = -\frac{p_{KE}}{\gamma} dK_X + \frac{t_Z}{\gamma} dZ.$$

Rewrite the above equation as:

$$p_{KE} dK_X = t_Z dZ - \gamma dE_X - \gamma dE_R. \qquad (D4)$$

Substituting equation (D4) to equation (D3) to find:

$$\frac{dU}{\lambda I} = \frac{1}{I} \frac{p_{KX}}{p_{KE}} (t_Z dZ - \gamma dE_X - \gamma dE_R) + \frac{p_{EX}}{I} dE_X + \frac{p_{ER}}{I} dE_R - \mu \frac{Z}{I} \hat{Z}$$

Rearranging the above equation shows:

$$\frac{dU}{\lambda I} = \underbrace{\left[\left(p_{EX} - \frac{p_{KX}}{p_{KE}}\gamma\right)\frac{E_X}{I}\hat{E}_X + \left(p_{ER} - \frac{p_{KX}}{p_{KE}}\gamma\right)\frac{E_R}{I}\hat{E}_R\right]}_{\text{Market Power}} + \underbrace{\left(\frac{p_{KX}}{p_{KE}}t_Z - \mu\right)\frac{Z}{I}\hat{Z}}_{\text{Externality}} \quad \text{(D5)}$$

Rearranging (D5) and substituting the definitions of $\varphi_R$ and $\varphi_X$ yields:

$$\frac{dU}{\lambda I} = \underbrace{\left(\varphi_X p_{EX} + \varphi_R p_{ER} - \frac{p_{KX}}{p_{KE}}\gamma\right)\frac{E}{I}\hat{E}}_{\text{Oligopoly Output}} + \underbrace{(p_{ER} - p_{EX})\frac{E_R E_X}{EI}(\hat{E}_R - \hat{E}_X)}_{\text{Price Discrimination}}$$

$$+ \underbrace{\left(\frac{p_{KX}}{p_{KE}}t_Z - \mu\right)\frac{Z}{I}\hat{Z}}_{\text{Externality}}. \quad \text{(D6)}$$

Assume pre-existing capital taxes are uniform in both industrial and energy sectors as $t_{KE} = t_{KX} \geq 0$, we will have $p_{KX} = p_{KE}$, and in that case equation (D5) becomes equation (15), and equation (D6) becomes equation (16) in Section 3.

Define $m_{EX} \equiv p_{EX} - \gamma - t_{EX}$ and $m_{ER} \equiv p_{ER} - \delta - \gamma - t_{ER}$, as the net revenue for industrial and residential energy sales, respectively, then rewrite equation (D6) as:

$$\frac{dU}{\lambda I} = \left(\varphi_X p_{EX} + \varphi_R p_{ER} - \gamma + \gamma - \frac{p_{KX}}{p_{KE}}\gamma\right)\frac{E}{I}\hat{E} + (t_{ER} + \delta - t_{EX})\frac{E_R E_X}{EI}(\hat{E}_R - \hat{E}_X)$$

$$+ (m_{ER} - m_{EX})\frac{E_R E_X}{EI}(\hat{E}_R - \hat{E}_X) - \left(\frac{t_{KE} - t_{KX}}{p_{KE}}t_Z\right)\frac{Z}{I}\hat{Z} + (t_Z - \mu)\frac{Z}{I}\hat{Z}.$$

Finally, rearranging the above equation finds:

$$\frac{dU}{\lambda I} = \underbrace{(\varphi_X p_{EX} + \varphi_R p_{ER} - \gamma)\frac{E}{I}\hat{E}}_{W1} + \underbrace{\left(\frac{t_{KE} - t_{KX}}{p_{KE}}\gamma\right)\frac{E}{I}\hat{E}}_{W2}$$

$$+ \underbrace{(m_{ER} - m_{EX})\frac{E_R E_X}{EI}(\hat{E}_R - \hat{E}_X)}_{W3} + \underbrace{(t_{ER} - t_{EX} + \delta)\frac{E_R E_X}{EI}(\hat{E}_R - \hat{E}_X)}_{W4} \quad \text{(D7)}$$

$$+ \underbrace{(t_Z - \mu)\frac{Z}{I}\hat{Z}}_{W5} + \underbrace{\left(\frac{t_{KX} - t_{KE}}{p_{KE}}t_Z\right)\frac{Z}{I}\hat{Z}}_{W6}.$$

This part of the appendix reports the disaggregated welfare effect that account for the initial tax distortions that are not independently reported in Tables 2, 3, and 4. The first term ($W1$) is the direct oligopoly output effect due to market power. Even in the perfect competitive case, W1 will not be zero as $p_{EX}$ and $p_{ER}$ need to cover taxes and the residential distribution cost. However, W3 will be zero in the perfect competition case as $m_{EX}$ and $m_{ER}$ are the per unit profits. The second term ($W2$) is a production distortion effect through pre-existing capital taxes. Thus, in the positive but differential capital tax case, $W1 + W2$ is a total market power effect. Similarly, the fifth term ($W5$) is the direct externality effect and the sixth term ($W6$) is externality correction effect through pre-existing capital tax distortion. Both terms $W2$ and $W6$ are zero when $t_{KE}$ equals $t_{KX}$ such that the intensive margin capital distortion across sectors is eliminated. The total amount of capital is fixed and thus the model does not contain an extensive margin distortion unlike Fullerton and Metcalf (2001). Meanwhile, the third term ($W3$) is the price discrimination effect as determined by the profit margins. The fourth term ($W4$) consists of two parts and it is an additional price discrimination effect induced through pre-existing commodity taxes (as distinct from the capital taxes) as well as price effect from the residential distribution cost. Terms ($W3$) and ($W4$) will cancel out given uniform prices. However, when $p_{ER}$ needs to be higher than $p_{EX}$ to cover the additional distribution cost ($\delta$), only term ($W3$) is going to be zero even in the perfect competition cases, as shown in Cases 1.8 and 4.8 in Table D3. The tables below record the disaggregated welfare effects from the numerical exercises in Section 5 differentiated by the six-term welfare equation given above.

**Table D1. Detailed Welfare Effects from Table 2 (in 2012$ million)**

| Case: | 1.0 | 2.0 | 3.0 | 4.0 |
|---|---|---|---|---|
| Total: | 3327.5 | 4280.1 | 4604.5 | 5631.1 |
| | | | | |
| W1: | -1492.9 | -865.2 | -931.3 | -1575.1 |
| W2: | 317.1 | 183.8 | 197.8 | 334.6 |
| W3: | 52.6 | 355.0 | 381.8 | 986.5 |
| W4: | 80.8 | 545.3 | 586.4 | 1515.3 |
| W5: | 4545.2 | 4224.2 | 4545.2 | 4545.2 |
| W6: | -175.4 | -163.0 | -175.4 | -175.4 |

(Notes: Cells are independent rounded.)

**Table D2. Detailed Welfare Effects from Table 3 (in 2012$ million)**

| Case: | 1.0 | 1.1 | 1.2 | 1.3 | 1.4 | 1.5 | 1.6 |
|---|---|---|---|---|---|---|---|
| Total: | 3327.5 | 2684.8 | 4287.7 | 910.1 | 6953.6 | 3482.5 | 3172.5 |
| | | | | | | | |
| W1: | -1492.9 | -1935.8 | -1437.1 | -1471.2 | -1525.5 | -714.3 | -2271.5 |
| W2: | 317.1 | 260.3 | 555.9 | 312.5 | 324.0 | 151.7 | 482.5 |
| W3: | 52.6 | 35.1 | 137.2 | 52.6 | 52.6 | 22.9 | 82.3 |
| W4: | 80.8 | 54.0 | 210.7 | 80.8 | 80.8 | 35.2 | 126.5 |
| W5: | 4545.2 | 4442.6 | 5014.5 | 2013.0 | 8343.6 | 4147.0 | 4943.4 |
| W6: | -175.4 | -171.4 | -193.5 | -77.7 | -322.0 | -160.0 | -190.8 |

(Notes: Cells are independent rounded.)

**Table D3. Detailed Welfare Effects from Table 4 (in 2012$ million)**

| Case: | 1.0 | 1.7 | 1.8 | 4.0 | 4.7 | 4.8 |
|---|---|---|---|---|---|---|
| Total: | 3327.5 | 856.1 | 4265.2 | 5631.1 | 22282.1 | 4967.8 |
| | | | | | | |
| W1: | -1492.9 | -13903.7 | -481.0 | -1575.1 | -15269.8 | 450.4 |
| W2: | 317.1 | 2953.2 | 315.7 | 334.6 | 3242.3 | -295.6 |
| W3: | 52.6 | 526.2 | 0.0 | 986.5 | 9150.1 | 0.0 |
| W4: | 80.8 | 808.3 | 123.7 | 1515.3 | 14055.0 | 1726.4 |
| W5: | 4545.2 | 10892.4 | 4479.7 | 4545.2 | 11549.0 | 3210.4 |
| W6: | -175.4 | -420.3 | -172.9 | -175.4 | -445.6 | -123.9 |

(Notes: Cells are independent rounded.)

## Appendix E: General Equilibrium Solutions

As described in Section 2, a 14-equation log-linear system defines the displacement model. This appendix solves the system for closed-form solutions. From equations (12) to (14), substitute $\hat{p}_{KE} = \hat{t}_{KE}$ and $\hat{p}_{KX} = \hat{t}_{KX}$ to replace $\hat{p}_{KE}$ and $\hat{p}_{KX}$ in equations (3), (5), (7), and (11). Next, subtracting equation (4) from equation (5) to reveal:

$$\hat{p}_X = (1 - \theta_E^X)\hat{t}_{KX} + \theta_E^X \hat{p}_{EX}. \tag{E1}$$

Subtracting (10) from equation (11), respectively, yields:

$$\hat{\gamma} = \frac{t_Z Z}{\gamma E}\hat{t}_Z + \frac{p_{KE} K_E}{\gamma E}\hat{t}_{KE} = \rho_Z^E \hat{t}_Z + (1 - \rho_Z^E)\hat{t}_{KE} \tag{S1}$$

where an equation denoted "S" is a solution to the system. Plugging the above equation into equations (8) and (9) to find:

$$\hat{p}_{EX} = \frac{n\varepsilon_{EX}}{1 + n\varepsilon_{EX}}\left(\frac{t_{EX}}{p_{EX}}\hat{t}_{EX} + \frac{p_{KE} K_E}{p_{EX} E}\hat{t}_{KE} + \frac{t_Z Z}{p_{EX} E}\hat{t}_Z\right), \tag{S2}$$

$$\hat{p}_{ER} = \frac{n\varepsilon_{ER}}{1 + n\varepsilon_{ER}} \left( \frac{t_{ER}}{p_{ER}} \hat{t}_{ER} + \frac{p_{KE}K_E}{p_{ER}E} \hat{t}_{KE} + \frac{t_Z Z}{p_{ER}E} \hat{t}_Z \right). \tag{S3}$$

Plugging equation (S2) into equation (A1) to yield:

$$\hat{p}_X = (1 - \theta_E^X)\hat{t}_{KX} + \theta_E^X \frac{n\varepsilon_{EX}}{1 + n\varepsilon_{EX}} \left( \frac{t_{EX}}{p_{EX}} \hat{t}_{EX} + \frac{p_{KE}K_E}{p_{EX}E} \hat{t}_{KE} + \frac{t_Z Z}{p_{EX}E} \hat{t}_Z \right). \tag{S4}$$

Substituting equation (4) to equation (2) and rearranging shows:

$$\hat{E}_R = (1 - \theta_E^X)\hat{R}_X + \theta_E^X \hat{E}_X - \sigma_U(\hat{p}_{ER} - \hat{p}_X). \tag{E2}$$

Substituting equation (3) to equation (E2) to eliminate $\hat{R}_X$ and rearranging shows:

$$\hat{E}_R - \hat{E}_X = -\sigma_X(1 - \theta_E^X)(\hat{t}_{KX} - \hat{p}_{EX}) - \sigma_U(\hat{p}_{ER} - \hat{p}_X). \tag{E3}$$

To eliminate $\hat{E}$, plugging equation (6) into (10) provides:

$$\varphi_X \hat{E}_X + \varphi_R \hat{E}_R = \frac{p_{KE}K_E}{\gamma E} \hat{R}_E + \frac{t_Z Z}{\gamma E} \hat{Z}. \tag{E4}$$

Then get rid of $\hat{Z}$, substituting equation (7) into (E4) shows:

$$\hat{E} = \varphi_X \hat{E}_X + \varphi_R \hat{E}_R = \hat{R}_E + \sigma_E \frac{t_Z Z}{\gamma E} (\hat{t}_{KE} - \hat{t}_Z), \tag{E5}$$

Substituting equation (E3) into (E5) to remove $\hat{E}_X$, and rearranging yields:

$$\begin{aligned} \hat{E}_R &= \hat{R}_E + \sigma_E \frac{t_Z Z}{\gamma E} (\hat{t}_{KE} - \hat{t}_Z) - \sigma_X \varphi_X (1 - \theta_E^X)(\hat{t}_{KX} - \hat{p}_{EX}) - \sigma_U \varphi_X (\hat{p}_{ER} - \hat{p}_X) \\ &= \hat{R}_X + \sigma_X \theta_E^X (\hat{t}_K - \hat{p}_{EX}) - \sigma_U (\hat{p}_{ER} - \hat{p}_X). \end{aligned} \tag{E6}$$

Similarly, to eliminate $\hat{E}_R$ by substituting equation (E3) into (E5) finds:

$$\begin{aligned} \hat{E}_X &= \hat{R}_E + \sigma_E \frac{t_Z Z}{\gamma E} (\hat{t}_{KE} - \hat{t}_Z) + \sigma_X \varphi_R (1 - \theta_E^X)(\hat{t}_{KX} - \hat{p}_{EX}) + \sigma_U \varphi_R (\hat{p}_{ER} - \hat{p}_X) \\ &= \hat{R}_X + \sigma_X (\hat{t}_K - \hat{p}_{EX}). \end{aligned} \tag{E7}$$

Substituting equations (4) and (E5) into equation (2) to remove $\hat{X}$ and $\hat{E}_R$, respectively, then using equation (3) to further eliminate $\hat{E}_X$ provides:

$$\hat{R}_X - \hat{R}_E = \sigma_E \frac{t_Z Z}{\gamma E} (\hat{t}_{KE} - \hat{t}_Z) - \sigma_X(\varphi_X + \varphi_R \theta_E^X)(\hat{t}_{KX} - \hat{p}_{EX}) + \sigma_U \varphi_R (\hat{p}_{ER} - \hat{p}_X). \tag{E8}$$

With only two unknowns ($\hat{R}_X$ and $\hat{R}_E$) left in equation (A8), remove $\hat{R}_E$ by substituting equation (1) to yield:

$$\hat{R}_X = \sigma_U \omega_E \varphi_R (\hat{p}_{ER} - \hat{p}_X) - \sigma_X \omega_E (\varphi_X + \varphi_R \theta_E^X)(\hat{t}_{KX} - \hat{p}_{EX}) + \sigma_E \omega_E \frac{t_Z Z}{\gamma E} (\hat{t}_{KE} - \hat{t}_Z). \tag{S5}$$

Rearranging equation (1) recovers $\hat{R}_E = \frac{-(1-\omega_E)}{\omega_E} \hat{R}_X$ and therefore:

$$\hat{K}_E = -\sigma_U(1 - \omega_E)\varphi_R(\hat{p}_{ER} - \hat{p}_X) + \sigma_X(1 - \omega_E)(\varphi_X + \varphi_R\theta_E^X)(\hat{t}_{KX} - \hat{p}_{EX})$$
$$- \sigma_E(1 - \omega_E)\frac{t_Z Z}{\gamma E}(\hat{t}_{KE} - \hat{t}_Z). \tag{S6}$$

All the remaining input solutions are found by substituting equations (S5) and (S6) back to equations (E5), (E6) and (E7) revealing:

$$\hat{E} = -\sigma_U(1 - \omega_E)\varphi_R(\hat{p}_{ER} - \hat{p}_X) + \sigma_X(1 - \omega_E)(\varphi_X + \varphi_R\theta_E^X)(\hat{t}_{KX} - \hat{p}_{EX})$$
$$+ \sigma_E \omega_E \frac{t_Z Z}{\gamma E}(\hat{t}_{KE} - \hat{t}_Z), \tag{S7}$$

$$\hat{E}_R = -\sigma_U(1 - \omega_E\varphi_R)(\hat{p}_{ER} - \hat{p}_X) + \sigma_X[\theta_E^X - \omega_E(\varphi_X + \varphi_R\theta_E^X)](\hat{t}_{KX} - \hat{p}_{EX})$$
$$+ \sigma_E \omega_E \frac{t_Z Z}{\gamma E}(\hat{t}_{KE} - \hat{t}_Z), \tag{S8}$$

$$\hat{E}_X = \sigma_U \omega_E \varphi_R(\hat{p}_{ER} - \hat{p}_X) + \sigma_X[1 - \omega_E(\varphi_X + \varphi_R\theta_E^X)](\hat{t}_{KX} - \hat{p}_{EX})$$
$$+ \sigma_E \omega_E \frac{t_Z Z}{\gamma E}(\hat{t}_{KE} - \hat{t}_Z). \tag{S9}$$

Next, substituting equation (S9) to equation (2) shows:

$$\hat{X} = \sigma_U \omega_E \varphi_R(\hat{p}_{ER} - \hat{p}_X) + \sigma_X[\theta_E^X - \omega_E(\varphi_X + \varphi_R\theta_E^X)](\hat{t}_{KX} - \hat{p}_{EX})$$
$$+ \sigma_E \omega_E \frac{t_Z Z}{\gamma E}(\hat{t}_{KE} - \hat{t}_Z). \tag{S10}$$

Finally, substituting equations (S6) to equation (7) yields:

$$\hat{Z} = -\sigma_U(1 - \omega_E)\varphi_R(\hat{p}_{ER} - \hat{p}_X) + \sigma_X(1 - \omega_E)(\varphi_X + \varphi_R\theta_E^X)(\hat{t}_{KX} - \hat{p}_{EX})$$
$$+ \sigma_E\left[1 - (1 - \omega_E)\frac{t_Z Z}{\gamma E}\right](\hat{t}_{KE} - \hat{t}_Z). \tag{S11}$$

Define the relative price changes as follows: $\hat{A} \equiv \hat{p}_{ER} - \hat{p}_X$, $\hat{B} \equiv \hat{p}_{EX} - \hat{t}_{KX}$, and $\hat{C} \equiv \hat{t}_Z - \hat{t}_{KE}$, thus:

$$\hat{A} \equiv \hat{p}_{ER} - \hat{p}_X = \left(\frac{n\varepsilon_{ER}}{1 + n\varepsilon_{ER}}\right)\left(\frac{t_{ER}}{p_{ER}}\hat{t}_{ER} + \frac{p_{KE}K_E}{p_{ER}E}\hat{t}_{KE} + \frac{t_Z Z}{p_{ER}E}\hat{t}_Z\right)$$
$$- \theta_E^X\left(\frac{n\varepsilon_{EX}}{1 + n\varepsilon_{EX}}\right)\left(\frac{t_{ER}}{p_{ER}}\hat{t}_{ER} + \frac{p_{KE}K_E}{p_{ER}E}\hat{t}_{KE} + \frac{t_Z Z}{p_{ER}E}\hat{t}_Z\right) - (1 - \theta_E^X)\hat{t}_{KX},$$

$$\hat{B} \equiv \hat{p}_{EX} - \hat{t}_{KX} = \left(\frac{n\varepsilon_{EX}}{1 + n\varepsilon_{EX}}\right)\left(\frac{t_{EX}}{p_{EX}}\hat{t}_{EX} + \frac{p_{KE}K_E}{p_{EX}E}\hat{t}_{KE} + \frac{t_Z Z}{p_{EX}E}\hat{t}_Z\right) - \hat{t}_{KX},$$

$$\hat{C} \equiv \hat{t}_Z - \hat{t}_{KE}.$$

Then, the quantity solutions can be recursively defined as:

$$\hat{K}_X = \sigma_U \omega_E \varphi_R\hat{A} + \sigma_X \omega_E(\varphi_X + \varphi_R\theta_E^X)\,\hat{B} - \sigma_E \omega_E \rho_Z^E\hat{C}, \tag{S5'}$$

$$\hat{K}_E = -\sigma_U(1-\omega_E)\varphi_R\hat{A} - \sigma_X(1-\omega_E)(\varphi_X+\varphi_R\theta_E^X)\,\hat{B} + \sigma_E(1-\omega_E)\rho_Z^E\hat{C}, \qquad \text{(S6')}$$

$$\hat{E} = -\sigma_U(1-\omega_E)\varphi_R\hat{A} - \sigma_X(1-\omega_E)(\varphi_X+\varphi_R\theta_E^X)\,\hat{B} - \sigma_E\omega_E\rho_Z^E\hat{C}, \qquad \text{(S7')}$$

$$\hat{E}_R = -\sigma_U(1-\omega_E\varphi_R)\hat{A} - \sigma_X[\theta_E^X-\omega_E(\varphi_X+\varphi_R\theta_E^X)]\,\hat{B} - \sigma_E\omega_E\rho_Z^E\hat{C}, \qquad \text{(S8')}$$

$$\hat{E}_X = \sigma_U\omega_E\varphi_R\hat{A} - \sigma_X[1-\omega_E(\varphi_X+\varphi_R\theta_E^X)]\,\hat{B} - \sigma_E\omega_E\rho_Z^E\hat{C}, \qquad \text{(S9')}$$

$$\hat{X} = \sigma_U\omega_E\varphi_R\hat{A} - \sigma_X[\theta_E^X-\omega_E(\varphi_X+\varphi_R\theta_E^X)]\,\hat{B} - \sigma_E\omega_E\rho_Z^E\hat{C}, \qquad \text{(S10')}$$

$$\hat{Z} = -\sigma_U(1-\omega_E)\varphi_R\hat{A} - \sigma_X(1-\omega_E)(\varphi_X+\varphi_R\theta_E^X)\,\hat{B} - \sigma_E[1-(1-\omega_E)\rho_Z^E]\hat{C}. \qquad \text{(S11')}$$